\documentclass[preprint,pra,aps,showpacs,preprintnumbers,amsmath,amssymb,eqsecnum]{revtex4}

\usepackage{graphicx}
\input epsf.tex
\usepackage{subfig}
\usepackage{floatrow}
\def\eps{{\epsilon}}

\def\p{\mathbf{p}}

\def\ih{{ \frac{i}{\hbar} }}

\def\half{\frac{1}{2}}

\def\sz{\langle \sigma_z \rangle}

\def\sy{\langle \sigma_y \rangle}
\def\ria{{\rightarrow}}
\def\p2{{ \langle p^2 \rangle }}

\newcommand\beq{\begin{equation}}
\newcommand\eeq{\end{equation}}
\newcommand\bea{\begin{eqnarray}}
\newcommand\eea {\end{eqnarray}}

\begin{document}


\title{Classical Limit of the Quantum Zeno Effect by Environmental Decoherence}

\author{D.Bedingham}

\author{J.J.Halliwell}

\affiliation{Blackett Laboratory \\ Imperial College \\ London SW7
2BZ \\ UK }

\date{\today}

\begin{abstract}
We consider a point particle in one dimension initially confined to a finite spatial
region whose state is frequently monitored by projection operators onto that region.
In the limit of infinitely frequent monitoring, the state never escapes from the
region -- this is the Zeno effect. In the corresponding classical problem, by contrast,
the state diffuses out of the region with the frequent monitoring simply removing
probability. The aim of this paper is
to show how the Zeno effect disappears in the classical limit in this and
similar examples. We give a general argument showing that the Zeno effect is suppressed
in the presence of a decoherence mechanism which kills interference between histories.
We show how this works explicitly in two examples involving projections onto
a one-dimensional subspace and identify the key timescales for the process. We
extend this understanding to our main problem of interest,
the particle in a spatial region case, by coupling
it to a decohering environment. Smoothed projectors are required to give the problem
proper definition and this implies the existence of a momentum cut-off and minimum lengthscale. We show that the escape rate from the region approaches
the classically expected result, and hence the Zeno effect is suppressed, as long as the environmentally-induced fluctuations
in momentum are sufficiently large. We establish the timescale on which an arbitrary
initial state develops sufficiently large fluctuations to satisfy this condition.
We link our results to earlier work on the $\hbar \rightarrow 0 $ limit of the Zeno
effect. We illustrate our results by plotting the probability flux lines for the density matrix (which are equivalent to Bohm trajectories in the pure state case). These illustrate both the Zeno and anti-Zeno effects very clearly,
and their suppression.
Our results are closely related to our earlier paper (Phys. Rev. A88, 022128 (2013))
demonstrating the suppression of quantum-mechanical reflection by decoherence.


\end{abstract}

\maketitle

\section{Introduction}

The quantum Zeno effect is the fact that the state of a quantum system can remain confined to a given Hilbert subspace if monitored sufficiently rapidly. That is, under unitary evolution interspersed by projection operators, the probability of finding the state still in its initial subspace goes to $1$ as the time between projections goes to zero \cite{Zeno,Zeno2}. By contrast, in classical mechanics, no such effect exists -- frequent monitoring of a classical system simply removes probability but does not prevent the state's departure from a region of the classical phase space. The purpose of this paper is to investigate the relationship between these two facts and in particular, to determine the conditions under which
the quantum Zeno effect goes away in the classical limit.

We begin by describing the Zeno effect more precisely.
We consider a system initially confined to the Hilbert subspace ${\cal H}_S$
of the Hilbert space ${\cal H}$ characterized
by the projector $P$ onto ${\cal H}_S$, so its initial state $ \rho $, which may be pure or mixed, satisfies $P \rho P = \rho $.
We suppose that over a time interval
$\tau$, the system evolves unitarily but with projectors at time intervals
$\eps$ and there are $N$ projectors in all, so that $ \tau = N \eps $.
The survival probability, the probability that the system is still in ${\cal H}_S$ at the end of this
time interval, is then given by
\bea
p(\tau) &=& {\rm Tr} \left( P e^{ - \ih H \eps}  \cdots P e^{ - \ih H \eps}
P \rho  P e^{  \ih H \eps} P  \cdots  e^{  \ih H \eps} P \right)
\label{1.1}
\\
&=& {\rm Tr} \left( (P e^{ - \ih H \eps} P)^N \rho (P e^{  \ih H \eps} P)^N \right).
\eea
Introducing $\bar P = 1 - P$, we note that
\beq
P e^{ - \ih H \eps} P = P e^{ - \ih PHP \eps} \left( 1 - \frac{\eps^2}{2} P H \bar P H P \right),
\eeq
plus terms of order $\eps^3$, and it follows that
\beq
(P e^{ - \ih H \eps} P)^N = P e^{ - \ih PHP \tau} \left( 1 - \frac{\eps^2}{2} P H \bar P H P \right)^N,
\eeq
to the same order.
We thus find that the leading order behaviour of $p(\tau)$ for small $\eps$ is
\beq
p(\tau) = 1 - \frac {N \eps^2}{ t_z^2},
\eeq
where the Zeno timescale $t_z$ is given by
\beq
t_z = \frac{\hbar } { \langle H \bar P H \rangle^{\half}}
\label{zeno1}.
\eeq
For a pure initial state $| \psi \rangle$ and projections onto that state $P = |
\psi \rangle \langle \psi | $, $t_z$ is the familiar expression for
the Zeno time,
\beq
t_z = \frac {\hbar } {\Delta H }.
\label{zeno2}
\eeq
This familiar timescale is not defined in the case of infinite dimensional Hilbert
subspaces, the main case considered here, and alternative
expressions are then required.

It follows from these expressions that the survival probability
$p(\tau) \rightarrow 1 $ as $\eps \rightarrow 0 $ with $\tau $ held constant.
This is the Zeno effect.
The Zeno regime, in which the projections are sufficiently frequent to
significantly inhibit escape, is defined by the Zeno time $t_z$.
For an analogous classical system, by contrast, we would expect that $p(\tau) <1$
no matter how small $\eps$ is, and indeed, for an initial state with outgoing momenta,
$p(\tau)$ will go to zero for sufficiently large $\tau$, so there is complete escape.

We will investigate the suppression of the quantum Zeno effect
using the standard machinery of decoherence
through interaction with an environment, described by simple master equations of
Lindblad form \cite{Lin}. This mechanism has provided an understanding of the classical limit of many different
types of quantum effects
\cite{Har6,JoZ,Hal00,Zur1,Zur2,Zur3,GH2,GH1,HaDo}.
Our goal is to understand the precise way in which this mechanism works for the Zeno effect.

We will generalize the survival probability formula Eq.(\ref{1.1})
to the case of non-unitary evolution described by a master equation.
Our main results involve a particle in one dimension initially
localized to a spatial region $[-L/2,L/2]$ and frequently monitored by projections onto that region of the form
\beq
P = \int_{-L/2}^{L/2} dx |x \rangle \langle x |,
\label{proj}
\eeq
with decoherence produced by a standard environment of the type used in quantum Brownian
motion. We will also briefly consider some simpler models with projections onto
a one-dimensional subspace
to get some insight into the timescales and general mechanism involved.

We will give general arguments to show that the diagonalization produced by this
master equation evolution suppresses the Zeno effect in Eq.(\ref{1.1}) and restores the expected classical result. We will explore this process in detail and establish the regimes and timescales in which it operates

Our work has a connection to recent work of Facchi et al \cite{Fac}, who
computed the $\hbar \ria 0 $ limit of Eq.(\ref{1.1}), using the Wigner-Weyl
formalism \cite{Wig,KiNo}, thereby obtaining the expected classical limit. We do not take
this limit in our work. Rather, we in effect show that the fluctuations
produced by the environment create a regime which is similar in structure to
the $\hbar \ria 0 $ regime and therefore a similar story applies.

Strings of projectors of the type Eq.(\ref{proj}), which are onto infinite dimensional Hilbert subspaces,
are not on the face of it easy to analyze in the survival probability Eq.(\ref{1.1}).
However, the analysis is greatly facilitated
by the results of Refs.\cite{Ech,HaYe3}, which show that a string of $N$ projections $P$ onto $[-L/2,L/2]$ interspersed with unitary evolution is well-approximated by
evolution in the presence of a complex potential outside the region, $V_0 \bar P$,
where $\bar P = 1 - P$.
That is
\beq
P e^{ - \ih H \eps}  \cdots P e^{ - \ih H \eps} P
\approx P \exp \left( - \ih H \tau - \frac {V_0} { \hbar} \bar P \right),
\label{complex}
\eeq
where the relationship between $V_0$ and $\eps$ was derived
in Ref.\cite{HaYe3} and found to be
\beq
V_0 \approx \frac {\hbar} { \eps}.
\label{V0}
\eeq
Complex potentials of this type arise in many applications \cite{complex,HaYe1}.
This representation of the string of projectors
gives a useful alternative perspective on the Zeno effect and its classical limit.

Using this representation, it is easily shown that,
for $V_0 \ll E$, where $E = \langle p^2 \rangle / 2 m $ is the energy scale of the initial state,
the complex potential simply absorbs flux leaving the region
but for $V_0 \gg E$, the state is reflected back into the region, with total reflection
in the limit $V_0 \rightarrow \infty$. Hence the Zeno effect is seen as reflection
off the complex potential modeling the string of projectors.
Crucially, the reflection condition together with Eq.(\ref{V0}) indicates that, for a wave packet state, the Zeno effect
becomes important when the time $\eps$ between projections is sufficiently small
that
\beq
\eps \ll t_E,
\label{eps}
\eeq
where the timescale $t_E = \hbar / E$ is referred to as the energy time. This is the timescale that replaces Eq.(\ref{zeno2}) for this
type of model and will be a very important one for the present paper.
This timescale has arisen in a number of different places
\cite{HaYe1,Mar,AOPRU,GLM}. It is known, for example, to define a bound
on the precision of time measurements \cite{HaYe1,AOPRU}.

The energy time also arises from a simple inspection of Eq.(\ref{1.1}), especially
if written out explicitly in terms of propagators (as we will do below). From the
parameters in this expression, one can
easily construct the length scale $ ( \hbar \eps /m )^{\half}$ which is the
effective spatial grid size of the ``diffraction grating'' produced by the projections
at time spacing $ \eps$. Reflection occurs when the wavelength $\hbar / p$
of the incoming packet is larger than the grid size, that is,
\beq
\left( \frac { \hbar \eps } { m } \right)^{\half} \ll \frac { \hbar }{p}.
\label{grid}
\eeq
This is easily seen to be equivalent to Eq.(\ref{eps}).

Analogous results can be derived for more general initial states with a broad range of energies, by separating the state into wave packets peaked around given energy.
Furthermore, the condition Eq.(\ref{eps}) for reflection remains approximately valid for a mixed state with evolution according to a master equation \cite{BeHa}, which
is very relevant to the present work.

In this different perspective of the Zeno effect, in which the reflection off the
boundary of the region is emphasized, it is clear that the Zeno effect will be suppressed if there is a mechanism which pushes up the average energy
(or equivalently, the momentum). This is because, as one can see from Eq.(\ref{eps}),
reflection, and hence the Zeno effect, operates only on the low momentum parts of
the state and if the energy is pushed up to sufficiently high values, Eq.(\ref{eps})
is no longer valid and there is no reflection.

It is well-known that the coupling to the environment producing decoherence also
produces fluctuations. That is, the mechanism that diagonalizes the density matrix
also produces large momentum fluctuations and these will therefore cause Eq.(\ref{eps})
to fail and kill reflection. So although we have two different perspectives of the Zeno effect and its suppression,
one and the same mechanism works in both perspectives.
In most of this paper we will largely focus on the creation of large momentum
fluctuations.

The work described here is very closely related to our earlier paper
Ref.\cite{BeHa}, which
considered quantum-mechanical reflection off both real and complex potentials of
step function form $V_0 \theta (x)$, and sought to show that reflection is killed
in the presence of a decohering environment coupled to the particle.
There is experimental evidence for this effect \cite{ChTe}.
Because of the relationship between a complex potential and string of projection operators outlined above, this is therefore
equivalent to the problem of suppressing the Zeno effect for a string of projection
operators onto the region $x>0$. It was found, however, that in order to suppress
reflection, the environment has to be so strong that it quickly produces very large fluctuations in the incoming wave packet, so large that the distinction between
incoming and outgoing is completely blurred. This means that although the quantum-mechanical
process of reflection is killed there is still a significant flux of reflected momenta
due to the fluctuations, and only part of the incoming state actually escapes into
$x>0$.

We find that the same effects are also in evidence in the present model, but with
a key difference, namely that we are considering escape from a {\it finite} spatial
region $[-L/2,L/2]$, not a half-infinite one. This means that fluctuations in either
direction actually {\it enhance} the escape process, so that all of the initial
state eventually escapes, the correct classical limit. Note that it is natural in
the present model to consider a state with $ \langle p \rangle = 0$, which diffuses
out of the region, and this we will do. If the initial state has a non-zero average
momentum, it would be very similar to the case of reflection off a single step potential
already covered in Ref.\cite{BeHa}.
(We also note that
in Ref.\cite{BeHa},
a modified model in which the environment was coupled to the target, rather than
to the particle, gave a satisfactory account of the suppression of reflection without
the fluctuations problem.)

We begin in Section II by giving a very general argument, inspired by the decoherent
histories approach to quantum theory, as to why the diagonalizing mechanism producing decoherence also suppresses the Zeno effect. In Section III, we summarize the description
of environmental decoherence in terms of a master equation for the density matrix.
We also describe the associated Wigner function picture.

In Section IV, we consider the Zeno effect in the presence of a decohering environment in two simple models.
The first is a simple
spin system, in which the spin in the $z$-direction is monitored, with evolution
described by a Lindblad master equation. The second is a point particle model,
in which the monitoring consists of frequent projections onto a Gaussian pure state,
with the decohering environment of standard quantum Brownian motion type. In both
of these cases the projections are onto one-dimensional Hilbert subspaces and the
Zeno time Eq.(\ref{zeno2}) governs the Zeno effect and this is
killed if the timescale associated with the environment is sufficiently small in
comparison.

We turn to our main problem, the question of escape from a finite spatial region,
in Section V. We write down the survival probability formula Eq.(\ref{1.1}) for the
situation in which a decohering environment is present.
We also observe in Section V that the use
of exact projection operators such as Eq.(\ref{proj}) presents some difficulties
and it is therefore necessary to replace them with quasi-projectors similar to
Eq.(\ref{proj}) but with smoothed off edges.

In Section VI we examine in detail what happens to the
state as a result of each projection. The Wigner picture turns out to be a particularly
useful way to analyse this and we also make contact with the $\hbar \ria 0 $ limit
situation discussed by Facchi et al \cite{Fac}. We introduce the quasi-projectors necessary to make the problem well-defined. These have a smoothing width $a$,
a new physical parameter which leads to a momentum cut-off $\hbar/ a$, and
we show that this in turn may be fixed in terms of other parameters by equating it
to the problem's natural momentum cut-off.
We also consider the evolution of $ \langle p^2 \rangle $
under the action of the quasi-projectors and establish the condition under which the quantum effects of the quasi-projections are negligible. It is basically the
requirement that $ \p2$ is sufficiently large.

In Section VII, we give a complementary analysis using the complex potential description,
Eq.(\ref{complex}), with similar results. We obtain, in a different way, the condition on the momentum fluctuations
required to kill the Zeno effect.

In Section VIII, we give a detailed analysis of the timescales of the problem.
This section, more than any other, gives the most complete heuristic account of the
suppression of the Zeno effect and also establishes the timescale on which it is
killed.
This heuristic picture is significantly backed up
by our numerical results described in Section IX.

In Section X, we take a somewhat different tack and illustrate our results by plotting
the probability flux lines of the density matrix of our model. These are equivalent
to Bohm trajectories \cite{Bohm} for the pure state case. These show very clearly
the degree to which probability can flow out of the region, inhibited by the Zeno
effect and enhanced by the environment.
We summarize and conclude in Section XI.

\section{Why Decoherence Suppresses the Zeno Effect}

We first give a very general argument, inspired by the decoherent histories
approach to quantum theory \cite{GH2,GH1,Gri,Omn,Hal1,DoH},
indicating why the presence of a decoherence
mechanism is expected to suppress the Zeno effect.
We again consider a system initially confined to the Hilbert subspace characterized
by the projector $P$, so its initial state satisfies $P \rho P = \rho$.
Let us suppose that at some final time $t_f$,
the dynamics are such that the system has completely departed from its initial subspace,
so that
\beq
{\rm Tr} \left( P(t_f) \rho \right) = 0.
\eeq
where $P(t)$ denotes the usual Heisenberg picture projectors.
We now monitor the system at a series of earlier times $t_1 < t_2 < \cdots t_n <
t_f $, separated by intervals $\eps$,
to see if it is in the subspace at those times, using the class operator
\beq
C = P(t_n) \cdots P(t_1).
\eeq
The Zeno effect is the fact that, under progressively more frequent monitoring, the
probability of remaining in the initial subspace tends to $1$,
\beq
{\rm Tr} \left( P(t_f) C  \rho  C^{\dag} \right)\rightarrow 1.
\label{3.3}
\eeq

Introducing the negation of $C$, $\bar C = 1 - C $, we have
\bea
0 &=& {\rm Tr} \left( P(t_f) \rho \right)
\nonumber \\
&=& {\rm Tr} \left( P(t_f) C \rho C^{\dag} \right) + {\rm Tr} \left( P(t_f) \bar C \rho \bar C^{\dag} \right)
\nonumber \\
&+& {\rm Tr} \left( P(t_f) C \rho \bar C^{\dag} \right) + {\rm Tr} \left( P(t_f) \bar C \rho C^{\dag} \right).
\eea
The four terms on the right-hand side are the probabilities and off-diagonal elements of the decoherence
functional for the pair of histories $P(t_f) C$ and $ P(t_f) \bar C$. If there is
decoherence of histories, as there would be in the presence of a decohering environment,
then the off-diagonal terms would be zero,
\beq
{\rm Tr} \left( P(t_f) C \rho \bar C^{\dag} \right) = 0 = {\rm Tr} \left( P(t_f) \bar C \rho C^{\dag} \right),
\eeq
and we are left with
\bea
0 &=& {\rm Tr} \left( P(t_f) \rho \right)
\nonumber \\
&=& {\rm Tr} \left( P(t_f) C \rho C^{\dag} \right) + {\rm Tr} \left( P(t_f) \bar C \rho \bar C^{\dag} \right).
\eea
Since the last two terms are non-negative, this means that
\beq
{\rm Tr} \left( P(t_f) C \rho C^{\dag} \right) = 0,
\eeq
contrary to Eq.(\ref{3.3}).
Hence the Zeno effect is killed when there is
decoherence of histories.

Note that this is not a decoherent histories {\it interpretation} of the Zeno effect.
We are simply using the machinery of decoherent histories and in particular the well-known
fact that the off-diagonal terms of the decoherence functional are suppressed in
the presence of an environment.
Note also that there is a converse to the above result: the presence of the Zeno
effect prevents decoherence of histories, as has been noted by a number of authors
\cite{HaYePit}.

\section{The Zeno Effect in the Presence of an Environment}

We are interested in the suppression of the Zeno effect through environmental decoherence
so in this section we briefly set out the standard account of environmental decoherence
and its application to the Zeno effect.
We suppose that we have a system $S$
coupled to environment ${\cal E}$. The standard formula Eq.(\ref{1.1}) holds for
our combined system $S {\cal E}$, with $P$ denoting projections onto the system only.
Assuming, as is standard, a factored initial state, then the environment can be traced
out and, in the physical interesting regime in which the subsequent evolution is
Markovian, the evolution of the density operator for the system is described by
a master equation of Lindblad form \cite{Lin}
\beq
\dot \rho = - \ih [ H, \rho] - [ \mathcal{L}, [\mathcal{L}, \rho]].
\label{lin}
\eeq
Here $H$ denotes the system Hamiltonian and we are concentrating on the simple
case in which there is a single hermitian Lindblad operator $\mathcal{L}$, whose form
will be specified in the context of specific models. The solution to this equation
may be written in the form
\beq
\rho_t = J_0^t [ \rho],
\eeq
where $\rho$ denotes the initial system state. The standard expression Eq.(\ref{1.1})
may then be written
\beq
p( \tau) = {\rm Tr} \left( P J_0^\eps[\cdots P J_0^\eps [P \rho P] P \cdots ] \right).
\label{Jprob}
\eeq
This is the expression we need to evaluate, in various models, to determine the effects of the environment on the Zeno effect.

In the special case where the projectors are onto one-dimensional subspaces,  $P = | \psi \rangle \langle \psi | = \rho $, and Eq.(\ref{Jprob}) then factors into the
simple form
\bea
p(\tau )  &=& \langle \psi | J_0^\eps \left[ | \psi \rangle \langle \psi | \right] | \psi \rangle^N
\nonumber \\
&=& \langle \psi | \rho_\eps | \psi \rangle^N.
\label{K2}
\eea
In simple models such as those considered in the next section,
the effects of the environment on the Zeno effect therefore come down to examining
the form of Eq.(\ref{K2}) for small times. In the next section, we will look at two
models of this type. The first is a simple spin system for which $H$ and $\mathcal{L}$
will be specified.

In the second simple model, and more importantly, in the main model considered in this
paper -- escape from a spatial region -- the system
consists of a particle coupled to a thermal environment (quantum Brownian motion), described by a Lindblad equation with
$\mathcal{L} = (D / \hbar^2)^{1/2} \ \! \hat x$. In configuration space it has the form
\begin{align}
\frac{\partial \rho}{\partial t} = \frac{i \hbar }{2m}\left(\frac{\partial^2 \rho}{\partial x^2} - \frac{\partial^2 \rho}{\partial y^2}\right)
-\frac{D}{\hbar^2} (x-y)^2\rho,
\label{master0}
\end{align}
and its associated
density matrix propagator $J_0^t$ is well-known and is given by
\begin{align}
J(x,y,t | x_0,y_0,t_0)
=& \frac{m}{2\pi\hbar (t-t_0)} \exp\left( \frac{im}{2\hbar (t-t_0)}\left[(x-x_0)^2-(y-y_0)^2\right]\right)
                                \nonumber\\
                                &\times  \exp\left(- \frac{D(t-t_0)}{3\hbar^2}\left[ (x-y)^2 +(x-y)(x_0-y_0)+(x_0-y_0)^2  \right]\right).
\label{Jprop}
\end{align}
(See for example Refs.\cite{CaLe,AnHa}.)

The decoherence process described by Eq.(\ref{Jprop}) has a number of different timescales
associated with it. The first is the localization time
\beq
t_{loc} = \left( \frac { m \hbar } { D} \right)^{\half}.
\label{tloc}
\eeq
This is the timescale on which the second part of the exponential in Eq.(\ref{Jprop}),
describing decoherence, becomes comparable with the first part, describing unitary
evolution. It is independent of the initial state.
It is also the timescale on which the Wigner function becomes positive
when evolved with Eq.(\ref{Jprop}) \cite{DiKi}.


When the state has quantum coherence over a lengthscale $\ell$, another timescale
is also important, the decoherence time, $t_d = \hbar^2 / D \ell^2 $. This is the
timescale on which pure states evolve to mixed, as measured by $ {\rm Tr} \rho^2$,
for example and tends to be very short for superposition states
with large $\ell$ \cite{Zur2}.
This arises, for example, for a superposition of wave packets a distance $\ell$ apart,
one of the most important examples of a decoherence process.
However, the physical
situation considered in the present paper is rather different. The reflection process
entailed in the Zeno effect involves quantum coherences over very {\it short} lengthscales
and it is these that we need to suppress by decoherence. The corresponding decoherence
time for this process is therefore very long. This fact was encountered in our previous
work on suppression of reflection \cite{BeHa}.


The Wigner function picture of our problem also gives many insights into how the process
works \cite{Wig,KiNo}. It is defined by
\beq
W(p,X) = \frac {1 } {2 \pi \hbar} \int d \xi \ e^{ - \ih p \xi } \rho ( X+ \half
\xi, X - \half \xi ),
\eeq
and the master equation in terms of the Wigner function is
\beq
\frac {\partial W} {\partial t} = - \frac {p}{m} \frac {\partial W} {\partial x}
+ D \frac {\partial^2 W} {\partial p^2}.
\label{Wig0}
\eeq
It may be solved using propagator,
\beq
W(p,X,t) = \int dp_0 dX_0 \ K(p,X,t|p_0,X_0,t_0) \ W_0 (p_0, X_0),
\eeq
where $K(p,X,t|p_0,X_0,t_0)$ is defined by applying the Wigner transform to both
ends of the density matrix propagator Eq.(\ref{Jprop}). It is given in this case
by
\begin{equation}
K = N \exp \left( - \alpha (p - p_{cl} )^2 - \beta ( X - X_{cl}  )^2
- \epsilon ( p - p_{cl} ) ( X - X_{cl}  ) \right),
\label{K1}
\end{equation}
where $p_{cl}$ and $X_{cl}$ denote the classical evolution from
$p_0$, $x_0$ to time $t$,
\begin{equation}
p_{cl} = p_0, \quad X_{cl} = X_0 + \frac {p_0 (t-t_0) } {m}.
\end{equation}
The coefficients $\alpha$, $\beta$ and $\epsilon$ are given by
\begin{equation}
\alpha = \frac {1}{D(t-t_0)}, \quad \beta = \frac { 3 m^2} {D (t-t_0)^3}, \quad
\epsilon = - \frac{ 3 m} { D (t-t_0)^2}.
\label{K3}
\end{equation}
(See for example Refs.\cite{HaDo,AnHa}.)
It is clearly peaked about classical evolution between the initial and final phase
space points and indeed tends to $\delta$-functions on classical evolution as
$D \ria 0 $.

\section{Two Simple Models}

In this section we consider two simple models with one-dimensional projectors in which Eq.(\ref{K2}) can be evaluated exactly. These simple models give quick verification
of some of the ideas outlined above. However, the restriction to projections onto
one-dimensional subspaces is very different to the case of projections onto ranges
of position considered in our main model, as we shall see.

\subsection{Two state system}

We consider a simple two state system described by a Lindblad equation of the form
Eq.(\ref{lin}) with $H = \omega \sigma_x $ and the Lindblad operator $\mathcal{L}$ is taken
to be proportional to either $\sigma_x$ or $ \sigma_y $, where $\sigma_i$ denote the Pauli matrices
for $i =x,y,z$. We take both the projections $P$ and the initial state to be onto
spin up in the $z$-direction, $ P = | \!\uparrow \rangle \langle \uparrow \! | $, which may also be written
$P = (1 + \sigma_z)/ 2$. We evaluate Eq.(\ref{K2}) to determine the effects of the
environment.

We are interested in the probability of the system being found in the up state after evolving
for time $t$ under the Lindblad evolution Eq.(\ref{lin}). This is given by
\beq
\langle \ \uparrow \! | \rho_t | \! \uparrow \rangle  = \half \left( 1 + \sz_t \right).
\eeq
For the case $\mathcal{L} = D^{\half} \sigma_x $, we find from the master equation Eq.(\ref{lin})
that
\bea
\frac {d} {dt} \sy &=& - 2 \omega \sz - 4 D \sy,
\\
\frac {d} {dt} \sz &=& 2 \omega \sy - 4 D \sz.
\eea
These equations are easily solved and we thus find
\bea
\langle \uparrow | \rho_t | \uparrow \rangle &=& \half \left( 1 + e^{ - 4Dt}
\cos (2 \omega t) \right)
\nonumber \\
&=& 1 - 2 D t - (\omega^2 - 4 D^2 )t^2 + \cdots.
\eea
Inserted in Eq.(\ref{K2}) with $t= \eps$ we obtain
\beq
p (\tau) = 1 - 2 D \tau - (\omega^2 - 4 D^2 ) \eps \tau + \cdots.
\eeq
In the case $D=0$ we thus see the Zeno effect as $\eps \rightarrow 0 $ with Zeno
time $ 1 / \omega $. But for
any non-zero $D$, there is always a linear decay, no matter how small $\eps $ is.
Hence the presence of the environment restores the expected classical decay

The case $\mathcal{L} = D^{\half} \sigma_y$ is similar and yields the result
\beq
\langle \uparrow | \rho_t | \uparrow \rangle = \half \left( 1 + e^{ - 2Dt}
\cos (2 \sqrt{\omega^2 -D^2} t) \right).
\eeq
This has a similar expansion for small $t$ hence the same qualitative results.

\subsection{Point particle with projections onto a one-dimensional subspace}

Our second simple model is a point particle coupled to a thermal environment described
by the propagator Eq.(\ref{Jprop}). We take a pure initial state
\beq
\psi (x) = \frac{1} { (2 \pi \sigma^2)^{1/4} } \exp \left( - \frac {x^2} {4 \sigma^2}
\right),
\eeq
and the projections at each time are onto the same pure state. We again compute the
probability of finding the system in the same state after time $t$, which is
\bea
\langle \psi | \rho_t | \psi \rangle &=& \int dx dy dx_0 dy_0 \ J (x,y,t| x_0, y_0,
0 )
\nonumber \\
& \times & \frac{1} { 2 \pi \sigma^2} \exp \left( - \frac {1} { 4 \sigma^2} (x^2 + y^2 + x_0^2
+ y_0^2) \right).
\eea
The integrals can all be evaluated, at some length, with the result,
\beq
\langle \psi | \rho_t | \psi \rangle
= \left( 1 + \frac {t} {t_d} \right)^{- \half}
\left( 1 + \frac {t^2} {16 t_z^2} \left( 1 + \frac {4t} { 3 t_d} \right)
\right)^{-\half}.
\eeq
Here, we have introduced the Zeno time for this problem $t_z = m \sigma^2 / \hbar $ (which is
also the timescale on which the initial state spreads)
and the decoherence time $ t_d = \hbar^2 / ( D \sigma^2)
$. These are also related to the localization time since we have $t_d t_z = t_{loc}^2$.
For small $t$, we have
\beq
\langle \psi | \rho_t | \psi \rangle = 1 - \frac {t} {2 t_d}
- \frac {t^2} {32 t_z^2} + \cdots.
\eeq
Again we see that the Zeno behaviour is moderated by the appearance of a linear term
produced by the environment which does not vanish in the limit $ \eps \ria 0 $ when inserted in Eq.(\ref{K2}).
The linear term is present no matter how small $D$ is, so in this model also there will always be leakage
from the subspace, no matter how frequently the projections act.

\section{Escape from a Spatial Region -- Setting up the Problem}

We now begin the analysis of our main problem: escape from the spatial
region $[-L/2,L/2]$. This involves computing the escape probability Eq.(\ref{Jprob}),
which is not possible to do exactly so in the following sections we will
resort to approximations and numerical solutions. In this section we
outline what is involved in a direct approach and the problems that arise.

The escape probability formula Eq.(\ref{Jprob}) may be written out explicitly
\bea
p(\tau) &=& \int_{-L/2}^{L/2} dx_n \int_{-L/2}^{L/2} dy_n \int_{-L/2}^{L/2} dx_{n-1} \int_{-L/2}^{L/2} dy_{n-1} \cdots \int_{-L/2}^{L/2} dx_0 \int_{-L/2}^{L/2} dy_0 \ \delta (x_n - y_n)
\nonumber \\
& \times &
\prod_{k=1}^n J (x_k,y_k,t_k |x_{k-1},y_{k-1},t_{k-1} ) \rho_0 (x_0, y_0),
\label{ptau}
\eea
where the propagator is given by Eq.(\ref{Jprop}) and the $\delta$-function is
present because of the final trace. The times $t_k$ are in units of $\eps$,
$t_k = k \eps$.
Introducing the traditional variables
$X = \half (x+y)$ and $ \xi = x-y $,
the propagator is
\begin{align}
J(X,\xi,t | X_0,\xi_0,_0)
=& \frac{m}{2\pi\hbar (t-t_0)} \exp\left( \frac{im}{\hbar (t-t_0)}
(X-X_0) (\xi- \xi_0) \right)
                                \nonumber\\
                                &\times  \exp\left(- \frac{D(t-t_0)}{3\hbar^2}\left( \xi^2 +\xi \xi_0 +\xi_0^2  \right)\right),
\label{Jpropxi}
\end{align}
and the integration ranges become
\beq
\int_{-L/2}^{L/2} dx \int_{-L/2}^{L/2} dy = \int_{-L/2}^{L/2} dX \int_{-L + 2|X|}^{L - 2|X|} d \xi.
\eeq
As one can see from the form of the propagator, the decoherence process leads to very effective diagonalization so that all of the integrals become very strongly concentrated
around $ \xi_k = 0$. Many decoherence calculations therefore typically assume that
one may make the approximation
\beq
\int_{-L/2}^{L/2} dX \int_{-L + 2|X|}^{L - 2|X|} d \xi
\approx \int_{-L/2}^{L/2} dX \int_{-\infty}^{\infty} d \xi.
\label{intapprox}
\eeq
This approximation is problematic in our case but we return to this below.
If Eq.(\ref{intapprox}) holds, then
Eq.(\ref{ptau}) is conveniently rewritten in terms of the initial Wigner function
$ W(p,X)$ and the associated Wigner propagator $K(p,X,t|p_0,X_0,0)$, Eq.(\ref{K1}),
\bea
p(\tau) &\approx& \int_{-L/2}^{L/2} dX_n \int_{-\infty}^{\infty} dp_n \int_{-L/2}^{L/2} dX_{n-1}
\int_{-\infty}^{\infty} dp_{n-1}  \cdots \int_{-L/2}^{L/2} dX_0 \int_{-\infty}^{\infty} dp_0
\nonumber \\
& \times &
\prod_{k=1}^n K(p_k,X_k,t_k| p_{k-1},X_{k-1},t_{k-1} ) W(p_0,x_0).
\label{ptau2}
\eea
This equation describes classical stochastic dynamics of a Brownian particle
required to pass through a series of gates, implemented by the restricted integrals
over $X_k$, at time intervals $\eps$. These gates
simply remove probability -- there is no quantum-mechanical reflection and the
dynamics is not significantly changed by taking the limit $\eps \ria 0$ with $\tau
= n \eps $ fixed. Hence a system
described by these dynamics will undergo escape from the region. There is no Zeno
effect.

The above is a traditional argument used in decoherence studies, especially in the
decoherent histories approach to quantum theory, and is
a more concrete implementation of the argument given in Section II
that a decoherence mechanism should suppress the Zeno effect. However, as indicated,
considerable attention needs to be given to the approximation Eq.(\ref{intapprox})
in this case. It is a reasonable one in situations where the projections are infrequent,
so create little reflection,
the situation considered in most applications. But this approximation is clearly problematic for values of $X$ close to the boundaries $X = \pm  L/2$ and the Zeno effect
studied here depends very critically on behaviour in precisely this region since
this is where reflection effects we are trying to kill are generated.
This issue is resolved by replacing the exact projectors in the escape
probability Eq.(\ref{Jprob}) with smeared projectors and we describe this in detail
in the next section.

\section{Analysis of the Projection Process}

The escape probability Eq.(\ref{Jprob}) may be broken down into simple
timesteps consisting of the action of the projections,
\beq
\rho \ \ria \ \rho' = P \rho P,
\eeq
followed by evolution with the propagator $J$. (In the unitary case, this is essentially the
process of diffraction in time \cite{diff}.)
When repeated many times this leads to the Zeno effect, but its origins
may be seen by looking very carefully
at the action of the projections at each time step.
In this section we examine
this process.

\subsection{The projection process in the Wigner picture}

Recall that the projections Eq.(\ref{proj}) have the form $ P = f_L ( \hat x ) $, where $f_L(x)$
is a window function on $ [-L/2,L/2] $, so we have in position space
\beq
\rho (x,y) \ \ria \ \rho'(x,y) = f_L(x) f_L(y) \rho (x,y).
\eeq
This clearly implies that probability is removed by this process, since
\beq
\rho' (x,x) = f_L(x) \rho (x,x),
\eeq
(noting that $f_L^2 (x) = f_L(x) $) but there are also off-diagonal effects and
in particular the momenta are also affected by the projections and this is the source
of reflection and the Zeno effect. This is all seen most clearly in the Wigner picture,
where the Wigner function $W' (p,X)$ of $\rho'$ is given by
\beq
W' (p,X) = f_L(X) \int dp_0 \ \frac { \sin \left(  \frac {(L-2|X|)} {\hbar}  ( p - p_0 ) \right)}
{ (p - p_0) } \ W( p_0, X).
\label{WigP}
\eeq
This very useful formula shows a number of relevant things. We first note there is
the factor $f_L(X)$, describing the expected loss probability. But more importantly,
the momentum distribution
is also affected -- a significant spreading of the momentum is
created by the projection, with the spreading largest close to the boundaries
of the region, $X= \pm L/2$.
When this process is extended to a large number of projections at sufficiently frequent
intervals, the spreading
process accumulates and concentrates into reflection.

This spreading is clearly a non-classical effect. In the naive limit $\hbar \ria 0 $,
the sinc term in Eq.(\ref{WigP}) (where ${\rm sinc}(x) = \sin (x) / x)$)
approaches a $\delta$-function
\beq
\frac { \sin \left(  \frac { (L-2|X|)} {\hbar}  ( p - p_0 ) \right)}
{ (p - p_0) } \ \ria \ \delta ( p -p_0),
\label{delta}
\eeq
and Eq.(\ref{WigP}) reduces to an essentially classical process describing nothing
more than the removal of probability. This observation is the essence of the
proof by Facchi et al that the Zeno effect goes away in the $\hbar \ria 0 $ limit
\cite{Fac}.

Furthermore, the formula Eq.(\ref{delta}) gives a strong indication why the environment
will suppress the Zeno effect: the left-hand side of Eq.(\ref{delta}) could
be {\it approximated}
by a $\delta$-function in Eq.(\ref{WigP}) if the Wigner function $W(p_0,X)$ is sufficiently broad in momentum
and large momentum fluctuations is precisely what the environment produces.
However, there is also a difficulty here, which is that the momentum width of $W(p_0,X)$
needs to beat
the momentum width of the sinc term, which is $\hbar / (L-2|X|) $, so can be arbitrarily
large close to the boundary $X = \pm L/2 $.

A related difficulty is that $W'(p,X)$ falls off rather slowly at large $p$ and as a consequence $ \p2 $ is divergent. This is due to the sharp edges of the projectors which generate a large number of very high momentum modes, a well-known feature of
sharp projectors in problems of this type \cite{diff,MaSc1,Sok1}.
It is not necessarily a problem, but
in the analysis that follows, $\p2$ is a very convenient indicator of the size of
the fluctuations so needs to be finite.

\subsection{Smeared projectors and the momentum cut-off}

The problem with $\p2$ becoming infinite is resolved by noting that there is in fact
a natural momentum cut-off in the problem. Although the momentum distribution is
indeed rendered extremely broad after each projection, the fact that another projection
occurs a time $\eps $ later means that the parts of the state with very high momentum
will have left the spatial region by that time and may therefore be discarded from
our considerations. This means that it is natural to ignore all momenta greater than
some cut-off value defined by
\beq
p_c = \frac { m L } { \eps}.
\label{pc}
\eeq
This momentum cut-off is actually most easily implemented indirectly, by introducing
a spatial smearing of the projectors. This leads to better comparison with the numerical
results, which have a natural shortest length defined by the lattice size. It also
smooths out the behaviour at the edges of the region, avoiding the large fluctuations
there.

We therefore replace the exact projector $P = f_L(\hat x )$ with a quasi-projector $ g_L( \hat x)$, where $ g_L ( x)$ is a smeared
window fuction with curvature scale $a$, with $a \ll L$, given by
\bea
g_L (x) &=&  \int_{-L/2}^{L/2} d \bar x \ \frac{ 1} { (2 \pi a^2)^{1/2} } \exp \left(-\frac {(x- \bar x)^2}{2a^2}\right)
\nonumber \\
&=& \int_{-L/2}^{L/2} d \bar x \ \delta_a (x - \bar x),
\label{smP}
\eea
where $\delta_a$ is a regularized $\delta$-function.
This clearly approaches $f_L(x)$ as $ a \ria 0 $. In terms of the quasi-projector,
the Wigner function of the projected density matrix is
\bea
W' (p,X) &=&  \frac{ 1} { ( \pi a^2)^{1/2} }\int_{-L/2}^{L/2} d \bar X \exp \left( - \frac { (X - \bar X)^2 } {a^2} \right)
\int dp_0 \ \exp \left( - \frac {a^2} { \hbar^2} ( p -p_0)^2 \right)
\nonumber \\
& \times &
\ \frac { \sin \left(  \frac { (L-2|\bar X|)} {\hbar}  ( p - p_0 ) \right)}
{ (p - p_0) } \ W( p_0, X).
\label{WigP2}
\eea
We thus see that the smearing gives better fall-off at large $p$ so
that $\p2 $ is finite. In particular, momenta greater than $\hbar / a $ are exponentially
suppressed, so to impose the physically sensible momentum cut-off Eq.(\ref{pc}),
we choose $a$ to be
\beq
a = \frac { \hbar } { p_c} =  \frac { \hbar \eps } { m L}.
\label{adef}
\eeq
Note also that this implies a relationship with
the effective grid size associated with the projections, Eq.(\ref{grid}),
\beq
(a L)^{\half} =  \left( \frac { \hbar \eps } { m} \right)^\half.
\eeq
This grid size is therefore the geometric mean of $a$ and $L$.

Now that $\p2$ is made finite, we may assert that
the left-hand side of
Eq.(\ref{delta}) is approximately a $\delta$-function in Eq.(\ref{WigP2}),
and the effects of the projections is essentially classical,
as long as $\p2$ is large enough. A more detailed calculation is required to see
how large.


\subsection{The evolution of $\p2$ with smeared projectors}
\label{SP2}

We now calculate in more detail the change in $\p2$ that takes place as a result
of the action of the smeared projector $g_L (\hat x)$, Eq.(\ref{smP}). This could in principle
be carried out using Eq.(\ref{WigP2}) but it is easier to work directly in configuration
space. We assume an initial pure state $\psi (x)$. Generalization of what follows
to a mixed state is straightforward. Under the process $ \psi (x) \ria g_L(x) \psi
(x) $, the final $\p2$ is given by
\beq
\p2_f = \hbar^2 \int dx \left( g_L(x)^2 | \psi'(x) |^2 + g_L (x) g_L'(x)
\frac {\partial} { \partial x} | \psi (x) |^2 + g_L'(x)^2 | \psi (x) |^2 \right).
\label{p2}
\eeq
The first term, $\p2^{red}$, is the value of $\p2$ reduced only
by the removal of probability due to the projection. It is finite as $a \ria 0 $
and may then be written,
\beq
\p2^{red} = \hbar^2 \int_{-L/2}^{L/2} dx \ | \psi'(x) |^2.
\label{pred}
\eeq
If this process is part of a
long sequence of projections, the reduction at each step will be small so this term
will be comparable in size to $\p2$ just before the projection.

We estimate the remaining terms for small $a$. Note that
\beq
g_L'(x) = \delta_a (x+L/2) - \delta_a (x-L/2).
\eeq
The second term in Eq.(\ref{p2}) is finite as $a \ria 0 $, so we may take the limit
with the result
\beq
\hbar^2 \int dx \ g_L (x) g_L'(x)
\frac {\partial} { \partial x} | \psi (x) |^2
= \frac {\hbar^2} {2} \left[ \frac {\partial} { \partial x} | \psi (x) |^2 \right]_{x=-L/2}
- \frac{\hbar^2} {2} \left[ \frac {\partial} { \partial x} | \psi (x) |^2 \right]_{x=L/2}.
\label{p2q1}
\eeq
It is natural in this problem to assume the state is symmetric about the origin so
the two terms on the right-hand side will be the same. In the case of an
initial state uniform across this region, $| \psi (L/2)|^2$ will be of order $1/L$
and the above term could then be of order $ \hbar^2 / (aL) $.
For other states more localized in the centre of the region,
$\psi$ is smaller on the boundary and the above term would
be negligible.

In the third term in Eq.(\ref{p2}), we may take the limit $a \ria 0 $ in one of the
factors of $g_L'(x)$ only, with the result
\beq
\hbar^2 \int dx \ g_L'(x)^2 | \psi (x) |^2
= 2 \left( \delta_a (0) - \delta_a (L) \right) | \psi (L/2) |^2,
\eeq
where we again have assumed a symmetric state. The term $\delta_a (L)$ will go to
zero as $a \ria 0 $ and the value of $\delta_a (0)$ may be read off from Eq.(\ref{smP}),
so we obtain the behaviour
\beq
\hbar^2 \int dx \ g_L'(x)^2 | \psi (x) |^2 \ \sim \ \frac {{\hbar^2}} {a} | \psi (L/2) |^2,
\label{p2q2}
\eeq
for small $a$. Again for a uniform initial state, this term could be of order $\hbar^2 / (a L)$,
but will be much smaller for other initial states. The numerics shows, as we shall
see, that Eq.(\ref{p2q1}) is generally much smaller than Eq.(\ref{p2q2}), so will
be neglected.

In brief, as a result of the smeared projections, $\p2$ changes according to
\beq
\p2_f \sim \p2^{red} + \frac {{\hbar^2}} {a} | \psi (L/2) |^2.
\label{proj1}
\eeq
For the case of a mixed state, relevant to what follows, $ | \psi (L/2)|^2 $
is replaced by $\rho (L/2,L/2)$. Under the conditions described above, most importantly
if $\psi$ is not unusually small on the boundary, this approximates to
\beq
\p2_f \sim \p2 + \frac {\hbar^2} { a L }.
\label{proj2}
\eeq
This means that the quantum effects of the projections, which lead to reflection
and the Zeno effect, will be negligible as long as $\p2$ becomes sufficiently large
that
\beq
\p2 \gg \frac {\hbar^2} { a L }.
\label{p2ineq}
\eeq
This is a more precise statement of the conditions under which the condition Eq.(\ref{delta})
holds as an approximation.
A smaller bound on the right-hand side of Eq.(\ref{p2ineq}) may be appropiate
in some circumstances but this represents the worst case situation.
Inserting the value of $a$ defined by the cut-off, Eq.(\ref{adef}), it reads
\beq
\p2 \gg \frac { m \hbar } { \eps},
\eeq
which is equivalent
to the expected condition $ \eps \gg t_E$.

\subsection{The energy time from the Wigner function picture}

Finally, the Wigner representation of the projection process also gives an alternative
indication that the key timescale between projections for the Zeno effect is $ \hbar
/ E$. Under time evolution in the unitary case, the evolution of the Wigner function is obtained by simply shifting $X$ to $X - p t / m $. It is sufficient to use the
unsmeared version Eq.(\ref{WigP}) and we obtain
\bea
W' \left(p,X- \frac{pt}{m} \right) & =& f_L(X- \frac{pt}{m} ) \int dp_0 \ \frac { \sin \left(  \frac { (L-2X)} {\hbar}  ( p - p_0 ) + \frac {2pt} {m \hbar } (p-p_0) \right)}
{ (p - p_0) }
\nonumber \\
& \times & \ W\left( p_0, X - \frac{pt} {m} \right),
\label{WigPt}
\eea
where for convenience we have taken $X>0$.
If $t$ is sufficiently large, the sinc term again leads to a $\delta$-function
approximation similar to Eq.(\ref{delta}), as long as
\beq
t \gg \frac { m \hbar} { | p (p-p_0)| }.
\eeq
For momenta close to complete reflection, $ p \sim - p_0 $, this timescale
coincides with the energy time. This means that for times longer than the energy
time, the momentum spreading effect is negligible and the evolution is approximately
classical. This confirms that the projections need to be made at time intervals
less than the energy time for significant reflection to accumulate.

\section{Complex Potential Approach}

A very useful alternative approach to evaluating the escape probability Eq.(\ref{ptau})
is to use the relationship Eq.(\ref{complex}) between a string of projectors and a complex potential and we briefly consider this.
This connection  means that our problem is approximately
equivalent to solving the master equation Eq.(\ref{master0}) in the presence of a
complex potential $V = V_0 \bar P$ (so $V$ is non-zero outside the region $[-L/2,L/2]$).
This will give a complementary take on the analysis of a single projection described in the previous section.

However, as will become clear, it will be necessary to assume that $V$ is smeared
over a length scale $a$ at its edges, in the same way that we smeared the projections,
using Eq.(\ref{smP}).
We will assert that there is a relationship of the form Eq.(\ref{complex}) between strings of {\it smeared} projectors and a {\it smeared} complex potential. This has not been proved although
it is very plausible, from the details of the derivation of Eq.(\ref{complex}) given
in Ref. \cite{HaYe3},

With a smeared complex potential in place,  Eq.(\ref{ptau}) is equivalent to solving the
master equation
\begin{align}
\frac{\partial \rho}{\partial t} = \frac{i \hbar }{2m}\left(\frac{\partial^2 \rho}{\partial x^2} - \frac{\partial^2 \rho}{\partial y^2}\right) -  \frac{1}{\hbar} \left(V(x)+V(y)\right)\rho
-\frac{D}{\hbar^2} (x-y)^2\rho.
\label{master1}
\end{align}
The diagonalization produced by the final term in this equation means that it is
reasonable to make the approximation
\bea
V(x) + V(y) &=& V(X + \half \xi ) + V( X - \half X)
\nonumber \\
&=& 2 V(X) + \frac {1} {4} \xi^2 V''(X) + \cdots.
\eea
It is then most convenient to switch to the Wigner picture in which the master
equation has the form
\bea
\frac {\partial W} {\partial t} = &-& \frac {p}{m} \frac {\partial W} {\partial x}
- \frac {2}{\hbar} V(x) W + D \frac {\partial^2 W} {\partial p^2}
\nonumber \\
&+&  \frac {\hbar }{4} V^{\prime\prime} (x) \frac {\partial^2 W} { \partial p^2}
+\cdots,
\label{Wig1}
\eea
where the ellipses denote higher order quantum terms involving odd powers of $\hbar $ and even derivatives of $V$ and $W$.

The point now is that the momentum diffusion produced
by the environment term spreads out the Wigner function and as a result the higher
order quantum terms involving odd powers of $\hbar$ are strongly suppressed.
The Wigner function
then evolves according to a classical stochastic theory with an absorbing potential
for which there is no reflection, only absorbtion of particles leaving the region.
There is therefore no Zeno effect in this situation.

More precisely, the leading quantum term will be negligible as long as
\beq
\left| \hbar^2  V^{\prime\prime} (x) \frac {\partial^2 W} { \partial p^2}
\right| \ \ll \left|  V(x) W \right|.
\label{ineq}
\eeq
A loose estimate of the second derivative term is
\beq
\frac {\partial^2 W} { \partial p^2} \sim \frac {W} { \p2},
\eeq
and an analysis similar to that applied to Eq.(\ref{p2}) indicates that
$ V''(X)$, if averaged in the state $W$, will be of order $ V_0 / (aL)$.
We thus find that, with these loose estimates, the inequality Eq.(\ref{ineq})
will hold under the condition
\beq
\p2 \gg \frac {\hbar^2}{ aL},
\label{p2ineq2}
\eeq
which is the same as the inequality Eq.(\ref{p2ineq}). Hence the complex potential picture is compatible with the analysis of a single smeared projection.

The smearing of complex potential, like the smeared projectors, is also easily seen
to cut off momenta greater than $\hbar / a $.
In the presence of a smeared potential, the reflection probability for a single
smeared step to lowest order in $V_0$ is
of order
\beq
|\psi_{\rm ref} (p)|^2  \sim \left( \frac {V_0} {E} \right)^2
\exp\left(-\frac{4a^2 p^2}{\hbar^2}\right),
\label{Uref}
\eeq
excluding other irrelevant factors \cite{BeHa}. This means again that momenta
greater than $\hbar / a $ are suppressed. Again it is natural to equate this
cut-off with $p_c$ defined in Eq.(\ref{pc}) and Eq.(\ref{p2ineq2}) is then equivalent
to $V_0 \ll E$.
Hence the condition Eq.(\ref{p2ineq2}) under which the non-classical terms in the Wigner
equation are negligible, which arises from considering density matrix diagonalization,
is the same as the condition $V_0 \ll E $
under which quantum-mechanical reflection from the boundary is negligible.


\section{Escape from a Spatial Region -- Timescale Analysis}

We now draw together the results of the last three sections to give an overall heuristic picture of the way the Zeno effect is suppressed in the problem of escape from a spatial region, and we also compute the relevant timescales.
We will assume that the particle starts out in a pure initial state with $ \langle p \rangle =0 = \langle x \rangle $. This is natural to assume since in the case of states with non-zero $\langle p \rangle $, this is then the problem of a wave packet
encountering a single boundary which was essentially covered in Ref.\cite{BeHa}.
We assume the state has spatial width $\sigma \sim L$, and $ \langle p^2 \rangle \sim
\hbar^2 / L^2 $. Since $ a \ll L$, the initial momentum is clearly much less than
the momentum cut-off $p_c$. This state could be a Gaussian state but we do not restrict to this choice.
(A natural assumption, which we use in some of the numerical work, is that the initial state has evolved for a while before the environment starts to act, so that a Zeno effect is established.)

For an initial state of the above type the energy time is
\beq
t_E = \frac{ \hbar m } { \langle p^2 \rangle } = \frac {mL^2} { \hbar},
\eeq
which is clearly the same as the spreading time over lengthscale $L$. This is the initial timescale on which the state departs from the region in the quantum case. In order to get a Zeno effect in the absence of an environment, it is therefore necessary that the time $\eps$ between projections satisfies
\beq
\eps \ll t_E = \frac {mL^2} { \hbar}.
\eeq
This requirement is clearly satisfied since it is equivalent to  $\hbar / L \ll p_c$.

The main point now is that in the presence of an environment, $\p2 $ increases,
which causes $t_E$ to decrease, and the Zeno effect will then be killed when $t_E$
drops below $\eps$. This will be achieved when
\beq
\p2 \gg \frac { \hbar m } { \eps}.
\label{p2con}
\eeq
Or equivalently, the analysis of the previous two sections indicates that the Zeno
effect is killed when $ \p2 $ is sufficiently large that Eq.(\ref{p2ineq2}) holds,
and since $a$ is given by Eq.(\ref{adef}) we arrive at exactly the same
condition, Eq.(\ref{p2con}).

Consider now how $\p2$ grows with time from its initial value and comes to satisfy
Eq.(\ref{p2con}). We have shown that under the action of a
single smeared projection $\p2$ changes according to Eqs.(\ref{proj1}), (\ref{proj2}). If in addition
it is then evolved for time $t \le \eps $ according to the master equation, then it is easily shown (from the Wigner form Eq.(\ref{Wig0}), for example), that
$\p2$ increases by $ 2 Dt$. The total change in $\p2$ is therefore given by
\beq
\p2 \ria \p2^{red} + \frac {\hbar^2} { a L } + 2 D t.
\label{Dt}
\eeq
The quantum term $ \hbar^2 / (a L)$ may be replaced with something
significantly smaller if the state is very small on the boundary,
as discussed, but this
formula is the appropriate one for a general initial state for the first few
projections. It is difficult to estimate the precise form of $\p2$ over longer
timescales after numerous projections have acted, but it is clearly the case that
$\p2$ will have increased by at least $ 2 Dt$ after time $t$, with the effects
of the projections enhancing this growth. Numerical results in the following
section confirm that the growth of $\p2$ is dominated by the $2Dt$ term in the initial
stages.
This behaviour means that the key condition Eq.(\ref{p2con}) will come to
be satisfied after a timescale $\tau$ which may be written in the equivalent
ways
\beq
\tau = \frac { m \hbar } { D \eps } = \frac {t_{loc}^2 } {\eps} = \frac {\hbar^2}
{ D a L}.
\label{tau}
\eeq
This is the timescale on which the Zeno effect is killed from a general initial
state. Note that it has the form of a decoherence time, for decoherence over
a lengthscale $(aL)^{1/2}$, which, as we observed, is the same as the effective grid size Eq.(\ref{grid}).

To see how $\tau$ compares to other timescales,
we compare the size $\eps $ and $ t_{loc}$. The case $ t_{loc} \ll \eps $ is
somewhat trivial. It means that, after each projection, the quantum term
in Eq.(\ref{Dt}) becomes negligible after time $\eps$ so the system is completely
classicalized before the next projection happens.
More interesting is the
case of a weaker environment, $ \eps \ll t_{loc}$ and in this case
\beq
\tau \gg t_{loc},
\eeq
so it takes longer than the localization time for the Zeno effect to be suppressed.

It is also of interest to see what happens on much longer timescales. The system will become classicalized
and is described by a classical
phase space distribution function $w(p,x)$ which evolves according to
\beq
\frac {\partial w} {\partial t} = - \frac {p}{m} \frac {\partial w} {\partial x}
+ D \frac {\partial^2 w} {\partial p^2}
\label{classWig},
\eeq
subject to probability being removed at time intervals $\eps$ by multiplying by the
window function $f_L(x)$. (This can also be modeled by a classical absorbing potential
acting continually).
Numerical solutions, described later, indicate that most initial states settle down
to a stationary state of the form
\beq
w(p,x,t) = \exp ( - \lambda t) w_s (p,x).
\eeq
Given the lengthscale $L$ of the region, we may, via these equations identify for this stationary solution a stationary momentum scale
\beq
p_s = (mLD)^{1/3},
\eeq
and a timescale
\beq
\lambda^{-1} = \left( \frac {m^2 L^2 } {D} \right)^{1/3} = \frac { m L } { p_s}.
\label{decayTime}
\eeq
The timescale $\lambda^{-1}$ thus determines the classical rate of decay from the region and our parameters will be chosen so that it is the longest timescale in the whole problem. In particular, if we choose $ \eps \ll \lambda^{-1}$, this is
equivalent to the natural condition that $p_s \ll p_c$,  so that $p_c$ is always
the highest momentum scale.
Furthermore, for the stationary solution to be clearly in the classical regime, we will require that it has a phase space spread which is much greater than $\hbar$, which means that
\beq
L p_s \gg \hbar.
\label{phase}
\eeq

In the stationary regime $ \p2$ acquires its stationary value $p_s^2$ and
it is natural to define a final value of the energy time,
\beq
t_E^f = \frac { m \hbar } { p_s^2} = \frac{ \hbar m^{1/3} } {(LD)^{2/3}}.
\eeq
Given this, we can check whether reflection is suppressed in this regime.
Eq.(\ref{phase}) implies that $t_E^f \ll t_{loc} \ll  \lambda^{-1}$.
This then means that there are three possibilities in terms of the magnitude of $\eps$ in relation to $t_{loc}$ and $t_E^f$, namely
\bea
t_E^f  & \ll  & t_{loc} \ll \eps \ll \lambda^{-1},
\\
t_E^f  & \ll  & \eps \ll t_{loc} \ll \lambda^{-1},
\\
\eps & \ll & t_E^f \ll t_{loc} \ll \lambda^{-1}.
\eea
In the first two of these inequalities, we have $t_E^f \ll \eps$ so we expect no quantum-mechanical reflection. In the third inequality $ \eps \ll t_E^f$ so reflection is possible in this case. This means that the Wigner function has entered its stationary
regime but the regime is still a quantum one. These features are all confirmed in the numerical solutions described below.

Furthermore, the first of these inequalities is the trivial case $ t_{loc} \ll \eps$.
The most interesting one with no reflection is the second one, into which we insert
the important timescale $ \tau$, Eq.(\ref{tau}), and it now reads
\beq
t_E^f  \ll   \eps \ll t_{loc} \ll \tau \ll \lambda^{-1}.
\label{regime}
\eeq

In summary, suppression of the Zeno effect is possible if the parameters of the model
satisfy the restriction Eq.(\ref{regime}). The effect is suppressed from a typical initial state on the timescale $\tau$, Eq.(\ref{tau}).

\section{Numerical results}

In this section we outline some numerical results on the Zeno effect and its suppression which substantiate the heuristic arguments of Section VIII. We will deal exclusively with the case of a particle in one dimension whose state is frequently monitored by projection operators acting on a region $[-L/2,L/2]$.

\subsection{Specification of numerics}
\label{SNUM}
We use numerical methods to solve the master equation (\ref{master0}) for the density matrix $\rho_t(x,y)$ interspersed with projections at time spacing $\epsilon$. All the numerical results outlined on this section are performed on a lattice with $2^8$ spatial points for each of the variables $x$ and $y$. We choose units such that $\hbar = m = L = 1$. In these units the lattice spacing in the space direction is chosen to be $\eta = 0.02$. The lattice spacing in the time direction is $\Delta t = 0.001$. In most of the examples, the initial state is a pure state Gaussian wavefunction with width $\sigma = 0.1$, i.e.
\begin{align}
\rho_0(x,y) = \frac{1}{\sqrt{2\pi\sigma^2}}\exp\left( -\frac{(x^2+y^2)}{4\sigma^2} \right).
\label{rhoIn}
\end{align}
This state is symmetric and well confined within the projection region $[-L/2,L/2]$. Subsequent projections onto this region occur at time intervals $\epsilon$. We will typically use $\epsilon = 0.01$.

The lattice spacing $\eta$ acts as a length scale for the sharpness of the edges of the projector. Although $\eta$ does not have precisely the same definition as $a$ (see Eq.(\ref{smP})), they play similar roles in terms of implementing a momentum cut-off and they have the same order of magnitude. The lattice momentum cut-off is in fact $p_c = \pi\hbar/\eta = 157$. This is the same order of magnitude as the physical cut-off $mL/\epsilon = 100$.

\subsection{The evolution of $\langle p^2\rangle$}

In Section \ref{SP2} we calculated the changes which occur to $\langle p^2\rangle$ as a result of a single projection and used this to determine the conditions under which the Zeno effect becomes negligible and the system behaves classically. Here we examine the evolution of $\langle p^2\rangle$ for a sequence of projections and confirm these earlier intuitions. The effect of a projection at time $t$ can be written as
\begin{align}
\langle p^2\rangle_t \rightarrow \langle p^2\rangle_{t+}  = \langle p^2\rangle^{red}_t + \Delta_t +\Sigma_t,
\end{align}
where $\langle p^2\rangle^{red}$ is the value of $\langle p^2\rangle$ reduced by the removal of probability due the projection, Eq.(\ref{pred}); $\Delta$ is the first quantum contribution to $\langle p^2\rangle$ given by Eq.(\ref{p2q1}); and $\Sigma$ is the second quantum contribution to $\langle p^2\rangle$ given by Eq.(\ref{p2q2}). All terms are renormalized by dividing by ${\rm Tr} P\rho$ --- the state norm following the projection.

Between projections the environment generates momentum fluctuations and we have
\begin{align}
\frac{d}{dt}\langle p^2\rangle = 2D.
\end{align}
Therefore, immediately prior to the next projection we have
\begin{align}
\langle p^2\rangle_{t+\epsilon} = \langle p^2\rangle^{red}_t + \Delta_t +\Sigma_t +2D\epsilon.
\label{pbreak}
\end{align}
The cycle then repeats with new values of $\langle p^2\rangle^{red}$, $\Delta$, and $\Sigma$ being generated by the next projection based on the updated momentum distribution. The system behaves classically when $\Delta$ and $\Sigma$ are negligible.

\begin{figure}[htb]
        \centering
        \sidesubfloat[]{%
                \includegraphics[width=.45\textwidth]{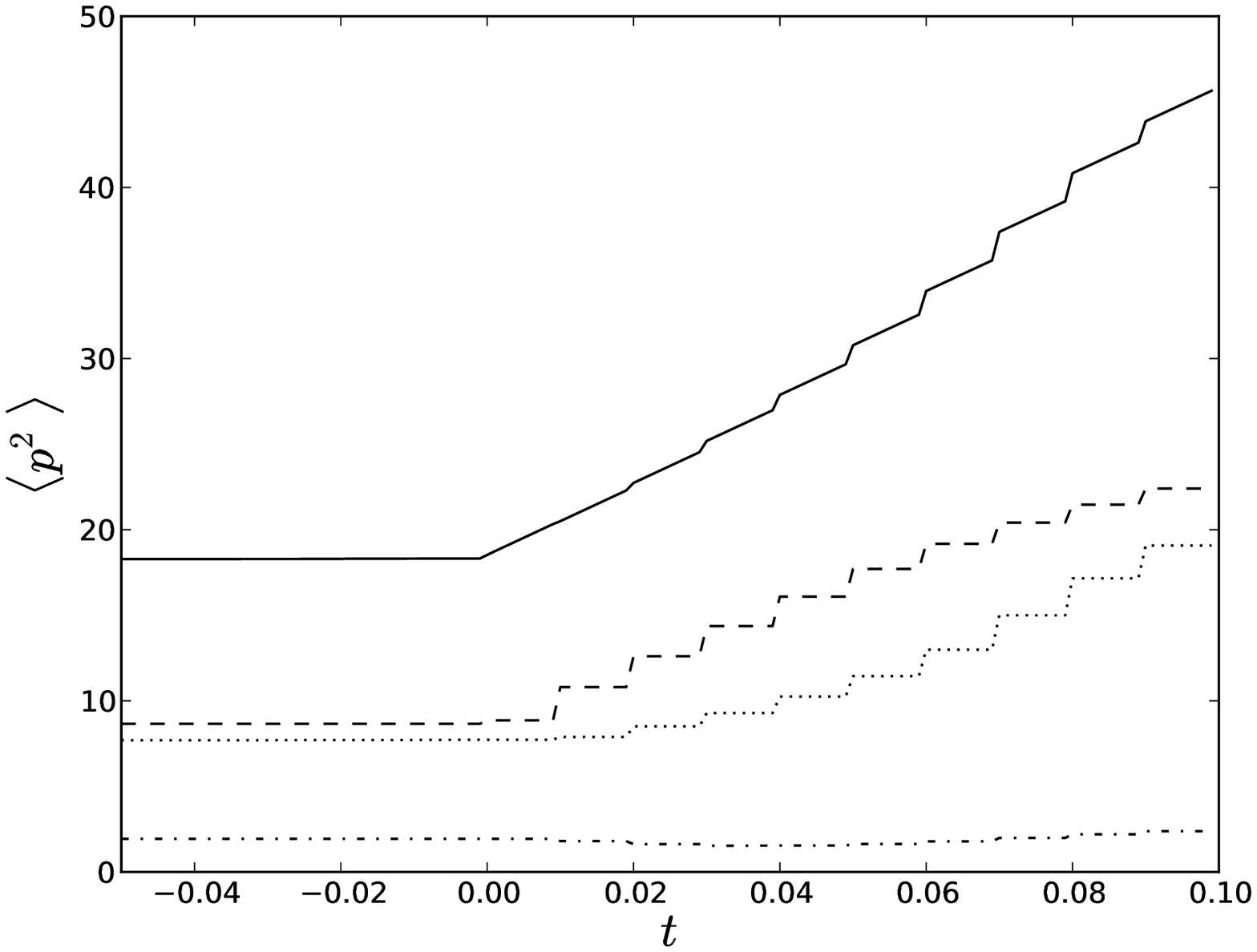}}
        \sidesubfloat[]{
                \includegraphics[width=.45\textwidth]{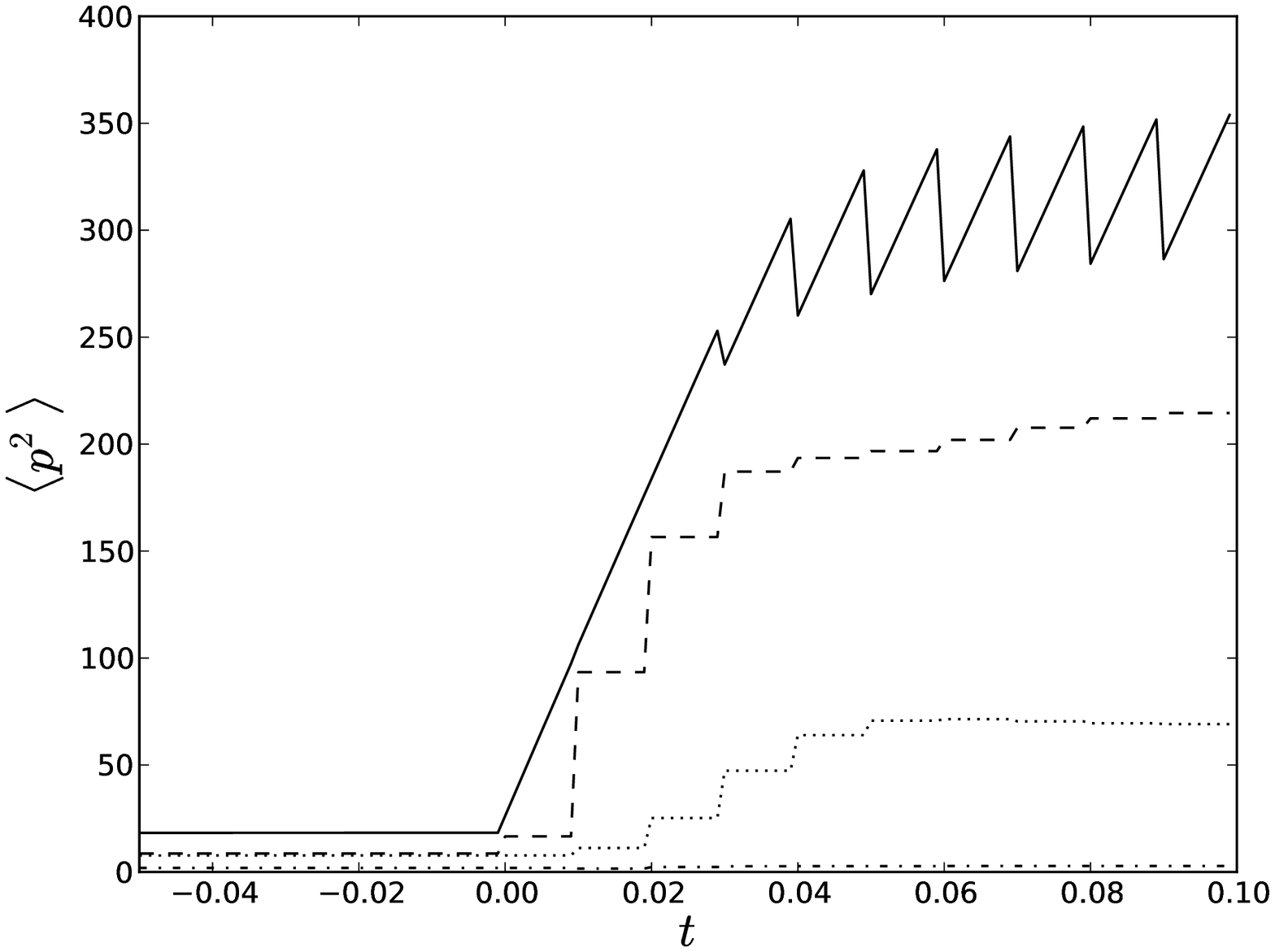}}\\
        \sidesubfloat[]{
                \includegraphics[width=.45\textwidth]{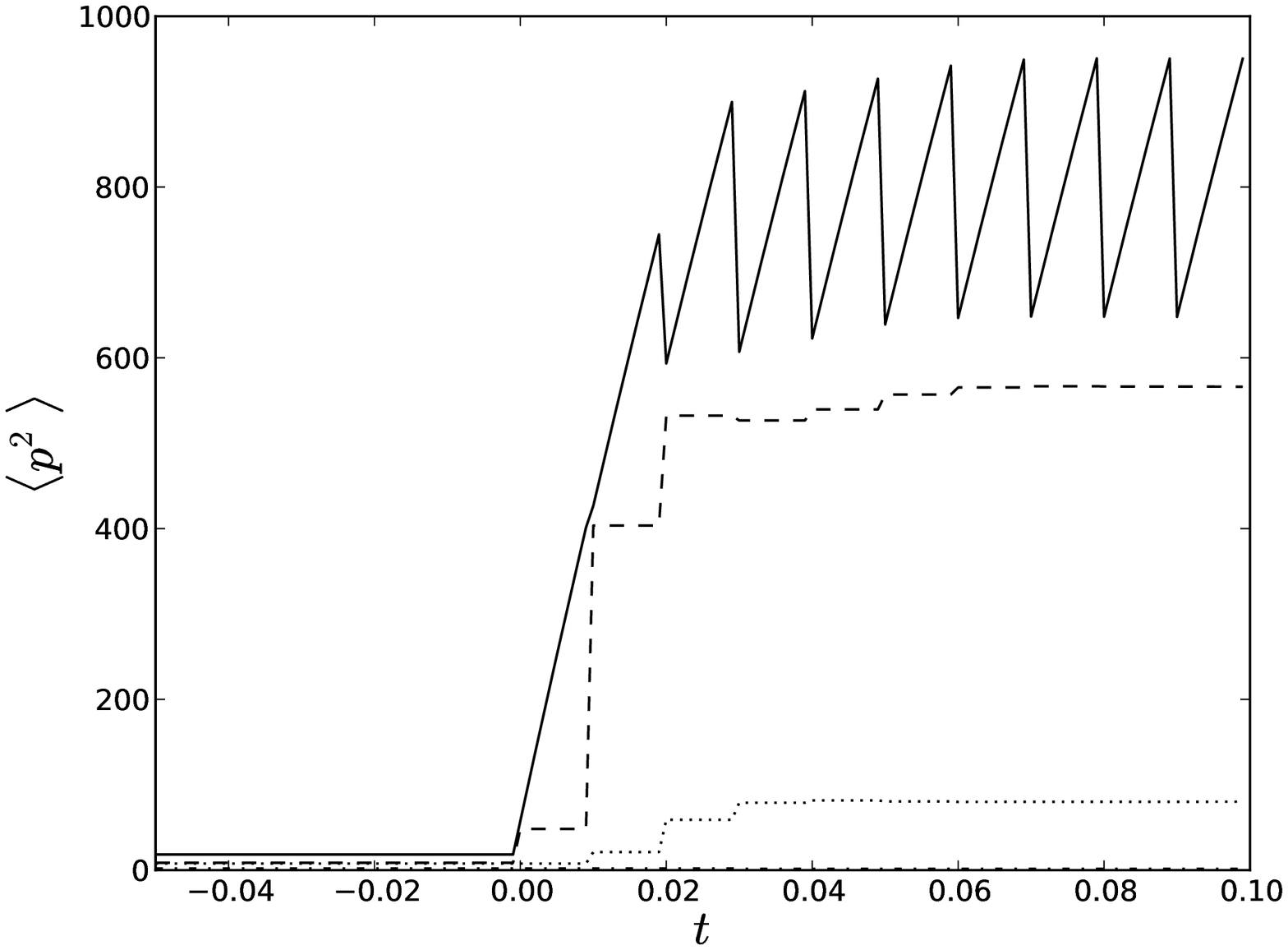}}
\caption{Evolution of $\langle p^2\rangle$ (solid line) for a sequence of projections with $\epsilon=0.01$. At $t=0$ an environment is introduced with (a) $D = 100$; (b) $D=4000$; (c) $D=20000$. Before $t=0$ the system has reached a steady state. $\langle p^2\rangle$ is broken down into (i) $\langle p^2\rangle^{red}$, Eq.(\ref{pred}) (dashed line); (ii) $\Delta$, Eq.(\ref{p2q1}) (dash-dotted line); (iii) $\Sigma$, Eq.(\ref{p2q2}) (dotted line).}
\label{Fp1}
\end{figure}

In the numerical computation of $\p2$, we have found that a number of different
initial states, when evolved without an environment, tend to settle down to
a stationary state. This is achieved when the increases in $\langle p^2\rangle$ due to quantum reflections from the boundary are offset by the removal of probability due to the projection. In this state the spatial probability distribution for the particle approximates a cosine peak (see Fig.~\ref{FX} below), although the state
is non-zero on the boundary.
We take this to be representative of a generic initial state.
In what follows, it is therefore natural to
start with the initial Gaussian Eq.(\ref{rhoIn}) at some initial time $t_0 \ll 0
$ and only switch on the environment at $t=0$, once the system is close to the stationary
state.

Figure \ref{Fp1} shows how $\langle p^2\rangle$ is affected by a sequence of projections after an environment is introduced at $t=0$. Environments of differing strengths are shown. In each case $\epsilon = 0.01$. Also shown are $\langle p^2\rangle^{red}$, $\Delta$, and $\Sigma$. These contributions to $\langle p^2\rangle$ are only updated after each projection. We find that between projections $\langle p^2\rangle$ increases at a rate $2D$ (once the environment is turned on) and the numerics confirm that Eq.(\ref{pbreak}) holds.
We also see that the quantum contribution
$\Sigma$ makes a large contribution to $\langle p^2\rangle$ at $t=0$ and therefore the quantum Zeno effect is not negligible initially. The other quantum contribution,
$\Delta$, generally makes only a small contribution.

In Fig.~\ref{Fp1}(a), after the introduction of the environment, $\langle p^2\rangle^{red}$ increases by approximately $2D\epsilon$ at each updating. The environment also leads to an increase in $\Sigma$. This is because $\rho (x,x)$ at the boundary increases as the spatial probability distribution becomes more uniform; the projections then result in a sharper discontinuity. For the parameters of this simulation it is not expected that a classical state will be reached since $t_E^f > \epsilon$. Indeed it is observed that the two contributions $\langle p^2\rangle^{red}$ and $\Sigma$ remain similar in size; the quantum contribution cannot be neglected.

In Fig.~\ref{Fp1}(b) the environment is significantly stronger. On dimensional grounds we expect $\Sigma\sim \hbar^2/\eta L = 50$ as an estimate of the size of the quantum contribution. This is confirmed by the numerics. After $t=0$, $\langle p^2\rangle$ is dominated by the $2Dt$ term and we can write $\langle p^2\rangle \simeq 2Dt$
as speculated in Section VIII. Furthermore, we can check in this figure the relevance of the Zeno suppressing timescale, Eq.(\ref{tau}). For these parameters, we have
$\tau = m\hbar/D\epsilon = 0.025$ and one can indeed see in the figure that
$\langle p^2\rangle$ becomes much larger than $\Sigma$ on precisely this timescale.
On longer timescales, the system eventually settles down to an almost classical steady state, again as expected from Section VIII and discussed below.

Figure \ref{Fp1}(c) shows the effect of an even stronger environment characterised by $D = 20000$. Here the Zeno suppression time is $\tau = 0.005$, which is very short and indeed the momentum variance $\langle p^2\rangle$ rapidly becomes much larger than $\Sigma$. By the time the system reaches a steady state, the quantum contribution (which remains of order $\hbar^2/\eta L$) makes a much smaller contribution to $\langle p^2\rangle$. The quantum Zeno effect is negligible.

\subsection{The steady state and the classical regime}

In the previous subsection we saw that the system settled into a steady state. In fact it is generally observed that in the presence of frequent projections onto the region $[-L/2,L/2]$, the system settles into a steady state  when increases in $\langle p^2\rangle$ due to the environment and the quantum Zeno effect are periodically offset by decreases due to removal of probability. In this section we will examine some features of the steady state and the conditions under which this state is approximately
classical.

\subsubsection{Evolution of the moments}

\begin{figure}[ht]
        \begin{center}
                \includegraphics[width=12.0cm]{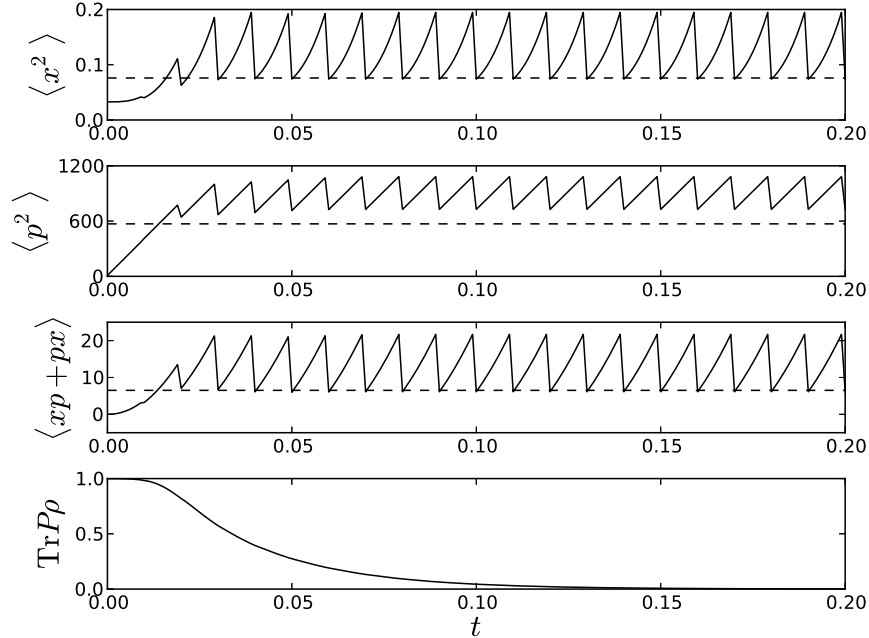}
        \end{center}
\caption{Time series of various moments. The solid lines correspond to an environment with diffusion parameter $D = 20000$. The dashed lines correspond to classical estimates for the steady state condition using the same value of $D$ with absorbing boundary conditions. The time spacing between projections is $\epsilon = 0.01$.}
\label{F7}
\end{figure}

Figure \ref{F7} shows a set of time series plots detailing the behaviour of the steady state. We display the moments $\langle x^2 \rangle$, $\langle p^2 \rangle$, $\langle xp+px \rangle$, and ${\rm Tr} P\rho$ where $P$ is the projection operator onto the spatial region. The plots use $D = 20000$ and the time spacing between projections is $\epsilon = 0.01$ in our units. Given the initial condition (\ref{rhoIn}) the steady state is achieved after a time of order $0.03$. Once in the steady state, in the times between the projections, the moments are seen to increase as the state escapes the projection boundary. The next projection restores the moments to the values they had after the previous projection. The plot of ${\rm Tr} P\rho$ indicates the probability for the particle to be found in the projection region.

\subsubsection{Comparison with the analogous classical system}

From Fig.~\ref{Fp1}(c) we see that with $D =20000$ and $\epsilon = 0.01$ the quantum Zeno effect can be neglected and that the system is behaving almost classically. In the classical regime we can treat the system in terms of a classical phase space distribution $w(p,x)$ satisfying (\ref{classWig}) and subject to a classical absorbing potential acting continually at $x = \pm L/2$. Equation (\ref{classWig}) can be written in a dimensionless form by expressing it in terms of parameters $\bar{x} = x/L$, $\bar{p} = p/p_s$, and $\bar{t} = t\lambda$. The resulting equation can be solved numerically with absorbing boundary conditions at $\bar{x} = \pm 0.5$ and we find a steady state solution with
\begin{align}
\langle \bar{x}^2 \rangle \sim 0.076 \quad ; \quad
\langle \bar{p}^2 \rangle \sim 0.78 \quad ; \quad
2\langle \bar{x}\bar{p} \rangle \sim 0.24.
\label{DBCM}
\end{align}
Rescaling for $D =20000$ results in a classical estimate of
\begin{align}
\langle {x}^2 \rangle \sim 0.076 \quad ; \quad
\langle {p}^2 \rangle \sim 570 \quad ; \quad
2\langle {x}{p} \rangle \sim 6.5.
\end{align}
These values are shown by the dashed lines in Fig.~\ref{F7}. We see that these values approximately correspond to the lower bound of the zigzag pattern, i.e.~the value of the moments immediately following a projection. The small discrepancy in $\langle p^2 \rangle$ is explained by the small quantum contribution $\Delta^b$ (see Fig.~\ref{Fp1}(c)). From Fig.~\ref{F7} we see that the half-life (i.e.~the time at which ${\rm Tr} P\rho = 0.5)$ is approximately 0.03. This corresponds well with the classical decay time $\lambda^{-1} = (m^2L^2/D)^{1/3}=0.037$ (Eq.(\ref{decayTime})).

\subsubsection{Wigner function of the steady state}

\begin{figure}[ht]
        \begin{center}
                \includegraphics[width=12.0cm]{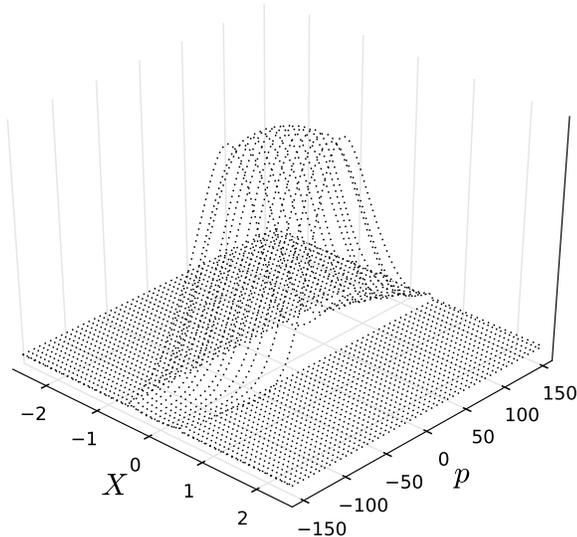}
        \end{center}
\caption{The Wigner function for the steady state with $D = 20000$.}
\label{F8}
\end{figure}

The form of the steady state is represented in Fig.~\ref{F8} which shows the Wigner distribution function in the case of $D=20000$ (cf.~Fig.~\ref{F7}). We see that the distribution is approximately uniform in position over the range $[-L/2,L/2]$, and approximately Gaussian in momentum. There is a skew in the distribution demonstrating the covariance between position and momentum.

\subsubsection{Spatial probability}

\begin{figure}[ht]
        \begin{center}
                \includegraphics[width=12.0cm]{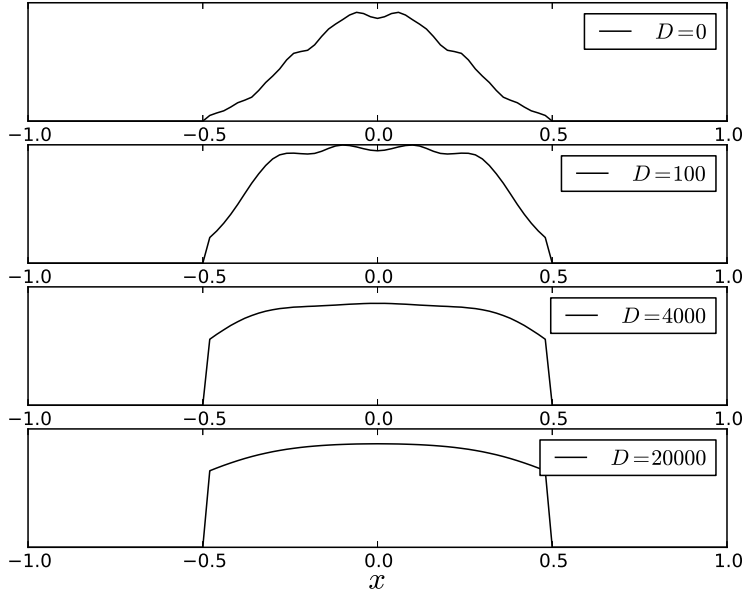}
        \end{center}
\caption{The spatial probability distribution $\rho(x,x)$ in the steady state condition immediately following a projection for different environments.}
\label{FX}
\end{figure}

Figure \ref{FX} shows the spatial probability distribution, $\rho(x,x)$, of the steady state immediately following a projection for a set of different environments. As we change $D$ from small to large values we see that the spatial probability distribution changes from a peaked form to a progressively more uniform distribution. We also note that for small $D$ we observe evenly spaced transient pulses on the probability distribution. These result from previous projections, reflecting probability back into the central region.

\subsubsection{The classical-quantum boundary in the steady state}

\begin{figure}[ht]
        \begin{center}
                \includegraphics[width=12.0cm]{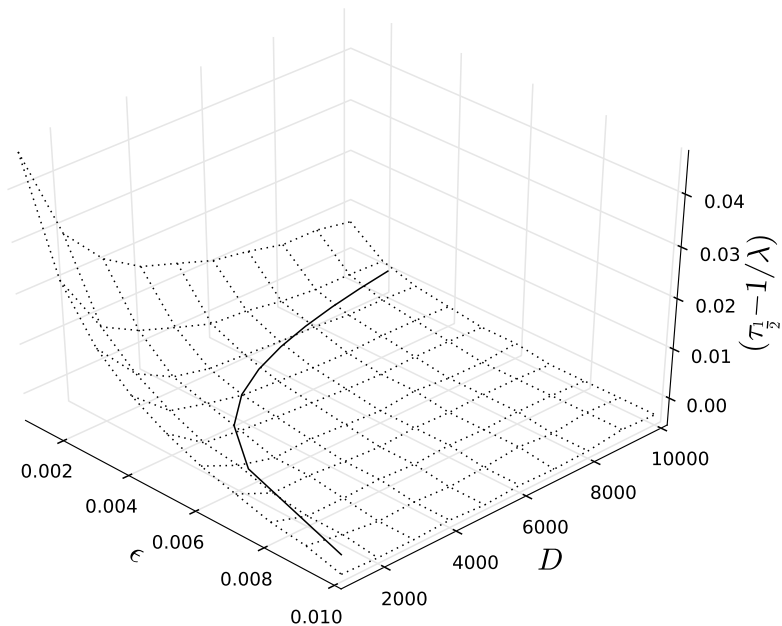}
        \end{center}
\caption{The half-life $\tau_{\frac{1}{2}}$ for the state to remain in the region [-0.5,0.5] minus the classical decay time $\lambda^{-1}$, with $D$ and $\epsilon$. The surface shows the numerical results for the full quantum system; the solid line indicates the theoretical condition dividing classical from quantum behaviour.}
\label{F10}
\end{figure}

The condition for the system to behave classically in the steady state is given by
\begin{align}
t_E^f = \frac{\hbar m^{1/3}}{(LD)^{2/3}} \ll \epsilon.
\label{DBCLASS}
\end{align}
In this regime we expect that the probability for the particle to remain in the projection region will decay on the classical decay time scale Eq.(\ref{decayTime}). We confirm this idea in Fig.~\ref{F10} which shows the difference between the half-life $\tau_{1/2}$ for the particle to remain in the projection region and the classical decay time $\lambda^{-1}$. This is shown for a range of $D$ and $\epsilon$ values. The solid line represents the condition $t_E^f = \epsilon$. To the right of this line we expect the system to behave classically. To the left we expect the quantum Zeno effect to appear. This is confirmed in a striking way. In the classical region $\tau_{1/2} \sim \lambda^{-1}$; in the quantum region $\tau_{1/2} > \lambda^{-1}$, i.e.~the system is confined to the projection region for longer than classically expected.

\subsubsection{Momentum variance and half-life in the steady state}

\begin{figure}[ht]
        \begin{center}
                \includegraphics[width=12.0cm]{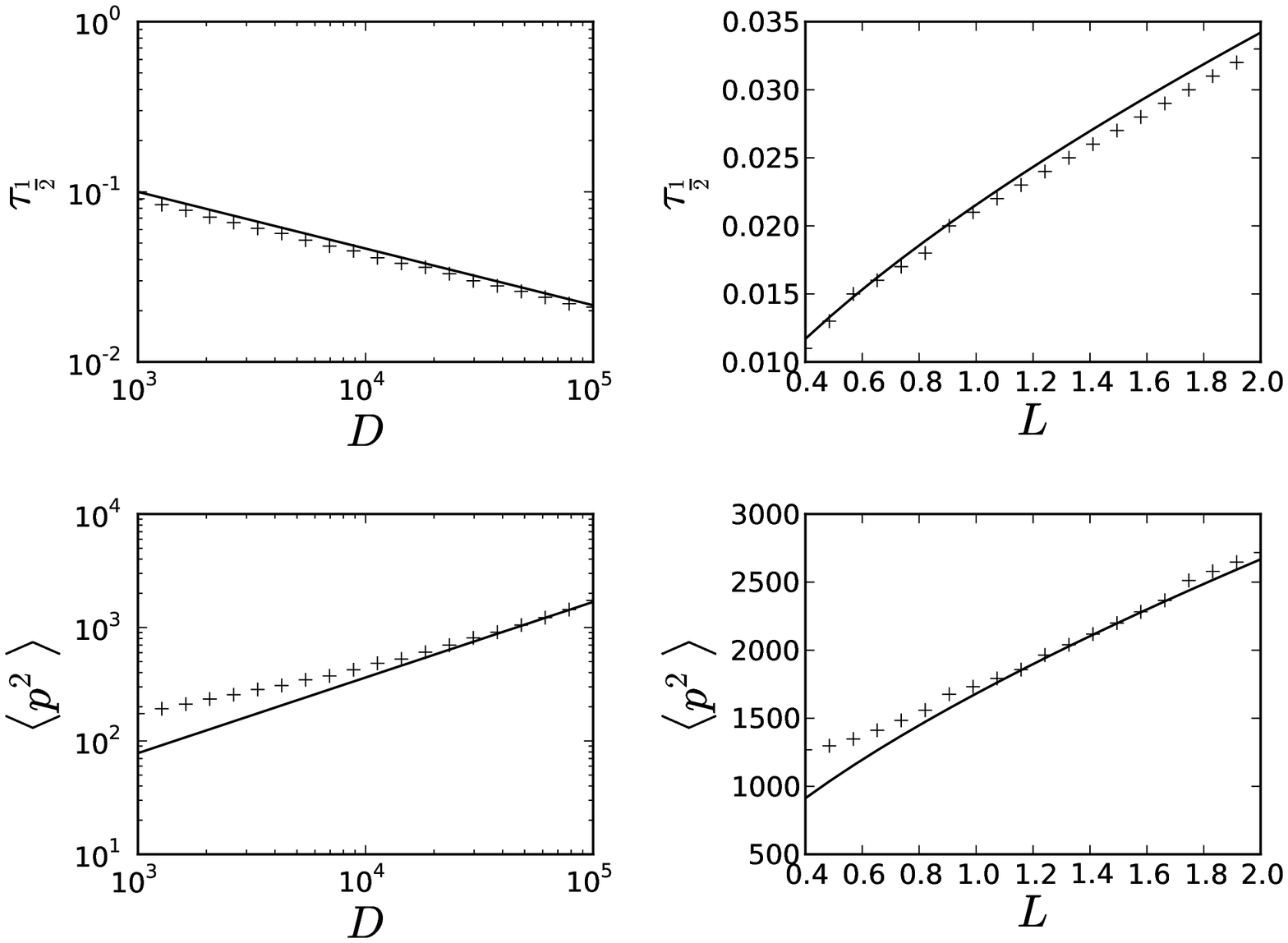}
        \end{center}
\caption{The behaviour of the half-life $\tau_{\frac{1}{2}}$ for the state to remain in the region $[-L/2,L/2]$ with $D$ (fixed $L=1.0$) and $L$ (fixed $D = 10^4$); and the variance of $p$, $\langle p^2 \rangle$ with $D$ (fixed $L=1.0$) and $L$ (fixed $D = 10^4$). The data points indicate numerical results for the full quantum system; the solid lines give the theoretical prediction assuming classical behaviour. The time spacing between projections is $\epsilon = 0.01$.}
\label{F9}
\end{figure}

In Fig.~\ref{F9} we confirm some of the predicted properties of the steady state in the classical regime. We expect the momentum variance to be given by
\begin{align}
\langle p^2\rangle = p_s^2 \langle \bar{p}^2 \rangle = (mLD)^{2/3} \langle \bar{p}^2 \rangle,
\end{align}
with $\langle \bar{p}^2 \rangle$ estimated in (\ref{DBCM}). We also expect the half-life $\tau_{1/2}$ to be given by the classical decay time $\lambda^{-1}$. These predictions are represented by the solid lines in Fig.~\ref{F9} for a range of values of $D$ (with $L$ fixed at $1.0$) and for a range of values of $L$ (with $D$ fixed at $10^4$). With a time spacing between projections $\epsilon = 0.01$, the full domains of the graphs in Fig.~\ref{F9} are expected to be in the classical regime according to (\ref{DBCLASS}). The data points are numerical results for the full quantum system. There is very good agreement between the theoretical and numerical estimates which improves as $D$ or $L$ increases.

\subsection{Summary of numerics}

In this section we have backed up some of the analytic conclusions of earlier sections with numerical evidence. We have examined the dynamical behaviour of $\langle p^2\rangle$ resulting from the combination of projections and interaction with an environment. We have distinguished the classical and quantum contributions to $\langle p^2\rangle$. In particular it was seen that for a sufficiently strong environment the initial behaviour of $\langle p^2\rangle$ following the introduction of an environment goes as $2Dt$. This confirms our heuristic argument in Section VIII that the Zeno suppressing timescale is  $\tau = m\hbar/D\epsilon$. After this time quantum contributions to $\langle p^2\rangle$ become relatively small and the system begins to behave classically.
We have also demonstrated the steady state behaviour and shown by comparison with the equations of motion for an equivalent classical phase space distribution (\ref{classWig}) that the system is behaving classically in this steady state provided that $t_E^f \ll \epsilon$. This was done by analysing the moments of the Wigner distribution once in the steady state, and their dependences on the parameters of the system.

\section{Illustration of the Zeno effect using probability flux lines}

In our main model of interest, involving escape from a spatial region, the Zeno effect and its suppression concern the containment or otherwise of probability in the spatial region $[-L/2,L/2]$. The effect is therefore very usefully illustrated by plotting probability flux lines showing the direction of flow of probability at each point in space-time. The velocity field of this flow at the point $x$ and time $t$ is defined by
\begin{align}
v_t(x) = \frac{J_t(x)}{\rho_t(x,x)},
\label{vel}
\end{align}
where
\begin{align}
J_t(x) = \frac{\hbar}{2mi}\left.\left[\frac{\partial}{\partial x}\rho_t(x,y)-\frac{\partial}{\partial y}\rho_t(x,y)\right]\right|_{x = y}.
\end{align}
A flux line is defined as a trajectory in this velocity field. Now note that the state satisfies the following continuity equation
\begin{align}
\frac{\partial}{\partial t}\rho_t(x,x) = - \frac{\partial}{\partial x}J_t(x) = - \frac{\partial}{\partial x}\left[\rho_t(x,x)v_t(x)\right],
\end{align}
where we have used Eq.(\ref{vel}). This implies that if the probability flux lines are initially distributed with a spatial density given by $\rho_0(x,x)$, they will continue to have a spatial density given by $\rho_t(x,x)$ for all future times $t$. The lines therefore provide a useful representation of the state, specifically they detail the spatial probability distribution for the location of the particle under consideration. Below we use this to illustrate the Zeno effect and to demonstrate the impact of an environment.

Note that if the density matrix represents a pure state, the probability flux lines correspond to Bohmian trajectories \cite{Bohm}. More generally for a mixed state the true Bohmian trajectories would be different for each pure contribution to the mixture and would not correspond to the probability flux lines as defined here.

We use numerical methods to first solve Eq.(\ref{master0}) for $\rho_t$ before determining the probability flux lines. The details of our simulations are the same as those outlined in Section \ref{SNUM}. We use the initial condition (\ref{rhoIn}) and units such that $\hbar = m = L = 1$.

\begin{figure}[htb]
        \centering
        \sidesubfloat[]{%
                \includegraphics[width=.45\textwidth]{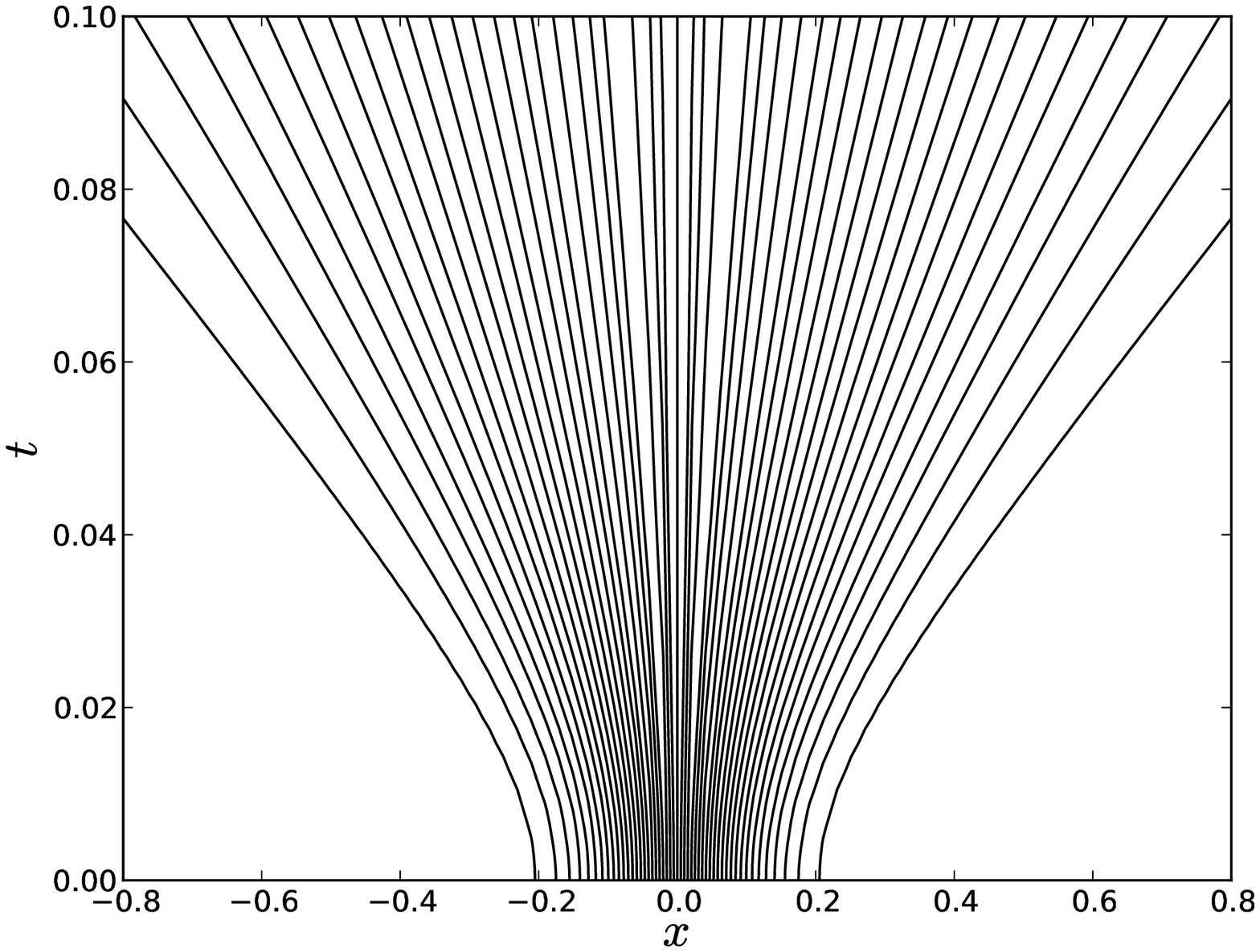}}\\
        \sidesubfloat[]{
                \includegraphics[width=.45\textwidth]{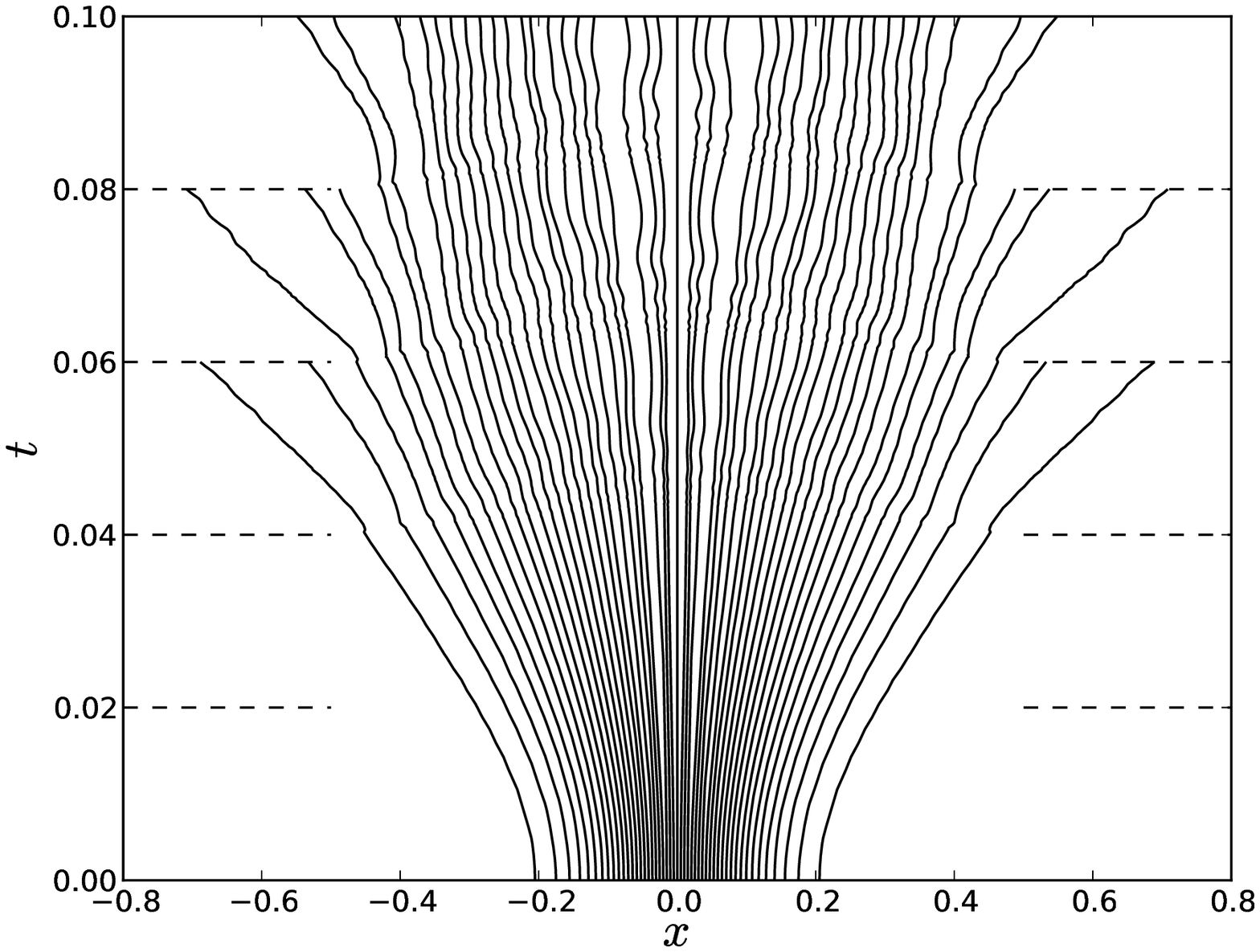}}
\caption{Probability flux lines for unitary evolution of a Gaussian wave packet: (a) no projections; (b) with a series of projections onto the spatial range $[-0.5,0.5]$ (marked with dashed lines). The initial Gaussian wavefunction in each case has width $\sigma = 0.1$ in units described in the text.}
\label{F1}
\end{figure}

In Fig.~\ref{F1}(a) we display the lines of probability flux in the case where there are no projections as a standard of comparison. As expected the lines spread out as the state undergoes standard quantum dispersion. The timescale associated with this spreading is $m\sigma^2/\hbar \sim 0.01$.

Figure \ref{F1}(b) shows the same system as Fig.~\ref{F1}(a) but with a sequence of projections occurring onto the spatial range $[-L/2,L/2]$ at time intervals $\epsilon = 0.02$. We see that the effect of a projection is to send a shock wave through the lines of probability flux which propagates in towards to centre of the wave packet. This has the effect of condensing those lines that are close to the centre, reducing the amount of spreading of flux lines with respect to the free case, Fig.~\ref{F1}(a), and helping to contain the probability within the projection region. This is the Zeno effect. Meanwhile, flux lines which are close to the boundary of the projection region are accelerated out of the projection region and absorbed by the next projection. This is the anti-Zeno effect. The probability flux plots reflect the behaviour of the wavefunction which is partly reflected off the boundary after a projection and partly spread out rapidly beyond the boundary in response to having a sharp discontinuity. The wiggles in the probability flux lines result from transient currents in the wavefunction generated by the sharp projections.

\begin{figure}[h]
        \centering
        \sidesubfloat[]{%
                \includegraphics[width=.45\textwidth]{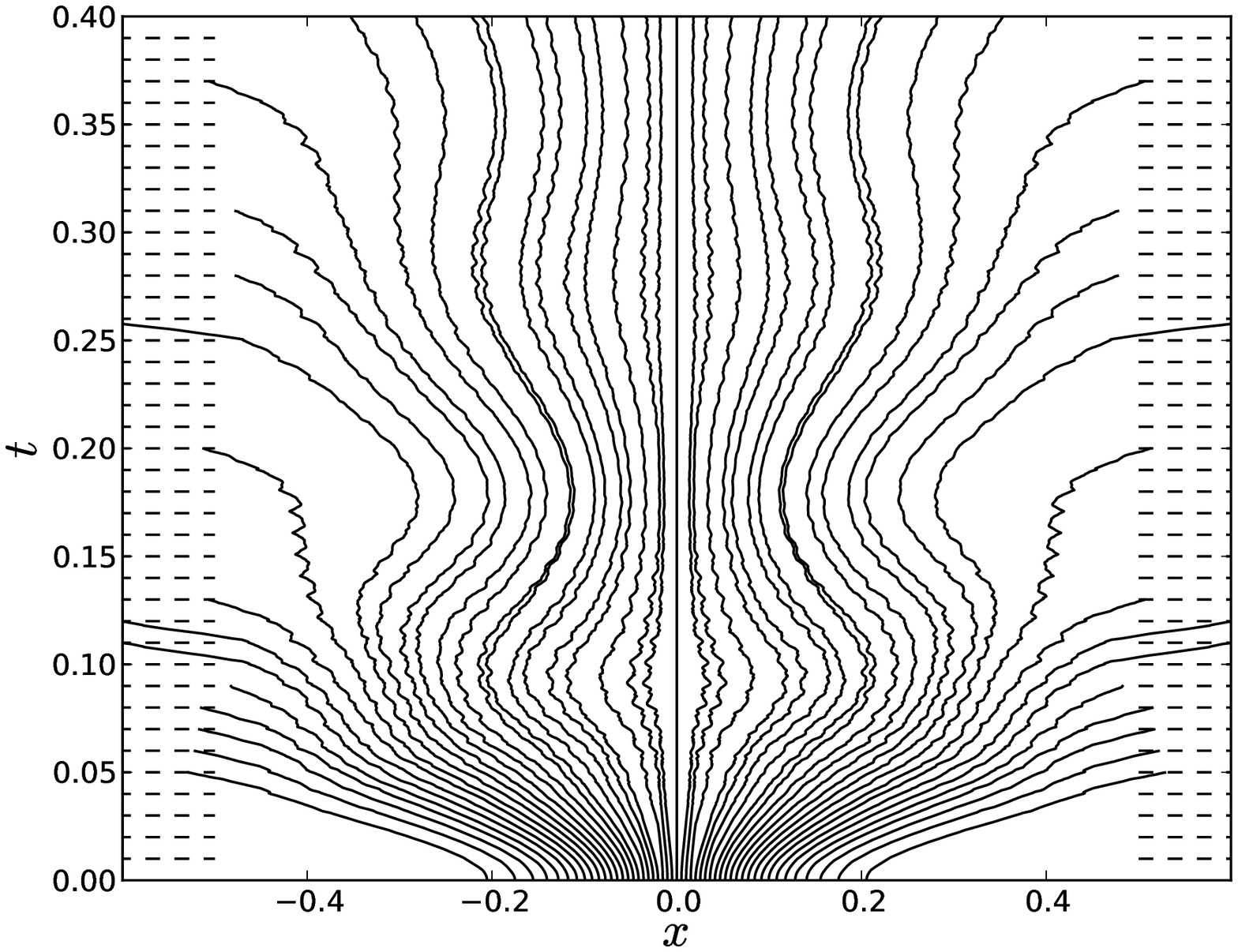}}\\
        \sidesubfloat[]{
                \includegraphics[width=.45\textwidth]{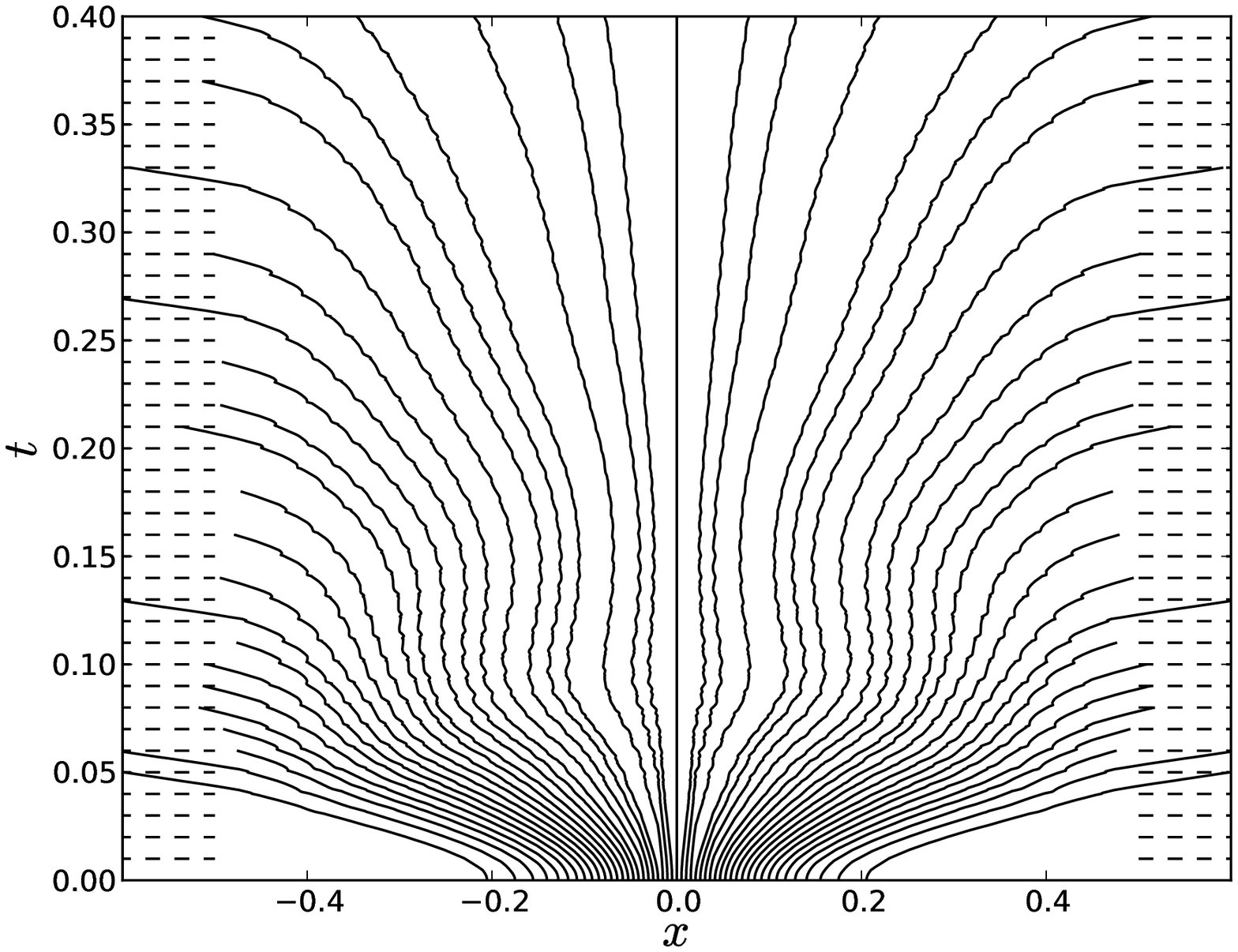}}\\
        \sidesubfloat[]{
                \includegraphics[width=.45\textwidth]{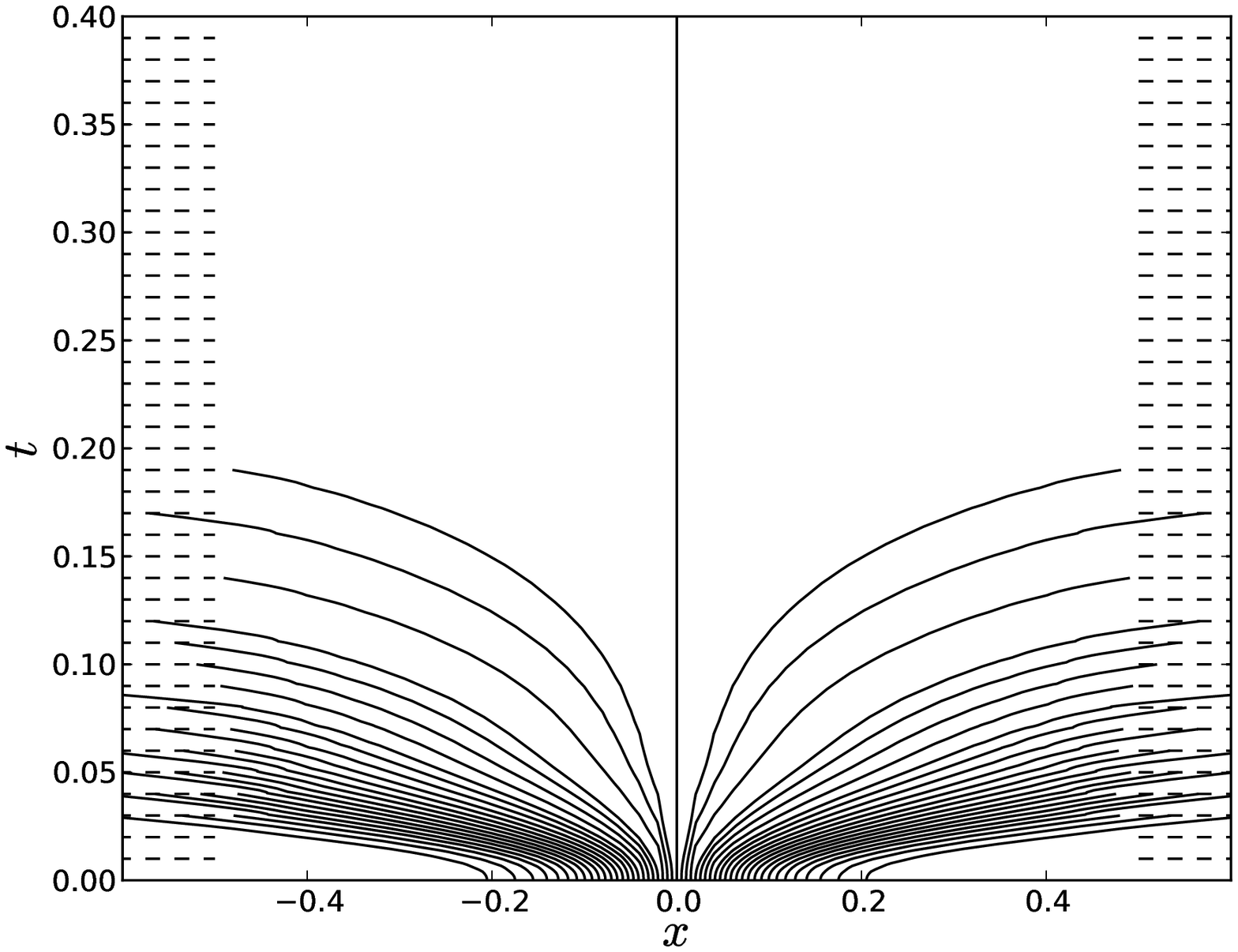}}
\caption{The effect of an environment on probability flux lines. Plots show evolution from an initial Gaussian wavefunction of width $\sigma = 0.1$ with (a) no environment; (b) environment with $D = 100$; (c) environment $D = 4000$. The time spacing between projections is $\epsilon = 0.01$ in units described in the text.}
\label{F3}
\end{figure}

Figure \ref{F3}(a) shows the result of an extended time period of frequent projections. Here the projections are spaced in time by $\epsilon = 0.01$. After an initial period in which the probability flux lines spread out through standard quantum dispersion, the lines appear to recondense in the centre at $t\sim 0.2$ before following the cycle once more. The period of this oscillation corresponds to the time taken for the initial wavefunction to spread to the lengthscale $L$ which is  $m\sigma L/\hbar = 0.1$. As the wavefunction reaches the boundary, part of it is reflected back towards the centre by the projections. This acts to contain the probability distribution for the particle in the projection region. If left for a longer period of time the oscillations settle down and the wavefunction tends approximately to the form $\cos(\pi x/L)$. This is the lowest energy solution for a wavefunction confined to the projection region. We also observe the anti-Zeno effect whereby those probability flux lines close to the boundary are rapidly accelerated outwards after a projection.

Figures \ref{F3}(b) and \ref{F3}(c) display the effect of an environment on this picture. The set up is the same as in Fig.~\ref{F3}(a). In Fig.~\ref{F3}(b) the environment in relatively weak; the diffusion parameter is set as $D = 100$ in the units described above. This does not cause significant change in the initial spreading of the probability flux lines, however, the effects of the projections are diminished. The flux lines are smoother than without an environment reflecting the fact that transient currents due to the projections are damped away by the environment. Although the flux lines do not condense to the same extent as in the case without an environment, there is still some containment of the flux lines in the centre. This indicates that the system is not behaving entirely classically and the Zeno effect is still important. The lower density of flux lines at later times in this figure indicates that the effect of the environment is to reduce the probability of the particle being found in the projection region with respect to the no environment case.

In Fig.~\ref{F3}(c) the diffusion parameter is $D = 4000$. Here the environment is sufficiently strong that the probability flux lines are visibly unaffected by the projections. The flux lines spread apart much more rapidly due to large momentum fluctuations resulting from interaction with the environment. There is no visible wiggle in the flux lines indicating that transient currents are eliminated by the environment. Also there is very little discernible deviation of the lines back towards the central region. (The flux line at $x = 0$ remains there due to the symmetry.) The characteristics of the Zeno effect (and the anti-Zeno effect) are not apparent in this figure; the system seems to be behaving in a classical way.

In summary, the environment leads to an increase in the rate at which the state disperses and a decrease in the quantum Zeno effect. Probability flux lines provide a useful way of illustrating the Zeno effect and its disappearance in the presence of a strong enough environment. They essentially show the probability distribution for the location of the particle and how it changes over a period of time.

\section{Summary and Conclusions}

The purpose of this paper was to study the suppression of the Zeno effect by environmental
decoherence, thereby establishing a sensible classical limit for systems exhibiting
this effect. The main system we focused on was a point particle in one dimension
monitored at time intervals $\eps$ by projecting onto the spatial region $[-L/2,L/2]$.
The Zeno effect arises when the projections are sufficiently rapid that $ \eps < \hbar / E $.
States trying to escape from the region are simply reflected from the boundary
under these conditions.
On the other hand, when $ \hbar / E < \eps $, the state is absorbed when it hits
the boundary so there is no Zeno effect.
The Zeno effect therefore exists only in the low momentum regime. It will be suppressed
if there is a mechanism which pushes the momenta to larger values.

On general grounds, environmental decoherence is expected to suppress all
quantum phenomena, but the way in which the Zeno effect is suppressed by decoherence
was not
immediately clear. We therefore gave, in Section II, an argument inspired by decoherent
histories showing that the Zeno effect would be absent in the presence of a decoherence
mechanism producing decoherence of histories. That is, density matrix diagonalization,
which is produced by environmental interactions, destroys the Zeno effect, in keeping
with the traditional understanding of emergent classicality.

Furthermore, since the variable $\xi = x-y$ in the density matrix is, via the Wigner transform, conjugate to momentum, density matrix diagonalization corresponds to the
development of large momentum fluctuations, precisely the desired mechanism for pushing up the momenta.
Our account of the suppression of the Zeno effect involves {\it two} different
but, as we showed, equivalent perspectives: density matrix diagonalization, the traditional story
of emergent classicality, and large momentum fluctuations required to suppress reflection.

The detailed story of how all this works was described in the remaining sections
of the paper. We described the mathematical techniques
of environmental decoherence in Section III and, in Section IV, showed how decoherence
kills the Zeno effect in two very simple models characterized by projections onto
one dimensional Hilbert subspaces.

The analysis of our main model commenced in Section V. Following a traditional
argument, we outlined the way in which the diagonalization mechanism of decoherence
turned the quantum expression for the survival probability into an essentially classical
one in which there is no reflection. However, we noted that this argument is problematic
in our case of very frequent projections,
essentially due to the potentially large contribution from reflection effects at
the boundary of the region.

In Section VI a detailed
analysis of the effects of each projection was carried out. The Wigner picture gave
a clear picture of the situation, showing the momentum spreading produced by the
position projections. We also noted the close connections with the
related results of Facchi et al, who considered the limit $\hbar \rightarrow 0 $ in
the unitary case.  We noted the improvements produced by smearing the projections,
the main improvement being that it induces a cut-off in the momenta, which
renders $\p2$ finite. We argued that this cut-off may be fixed in a physically sensible way.
We computed in detail the change produced in $\p2 $ by the projections, which is
key to understanding the Zeno effect. We also noted that the Wigner picture gave
another way of understanding the appearance of the energy time.
An alternative take on the analysis of Section VI was given in
Section VII, using the equivalent complex potential, with consistent results.

All of these results were drawn together into an overall heuristic analysis of the
suppression of the Zeno effect in Section VIII, which was backed up by numerical analysis
in Section IX. We showed that, in the appropriate regime of sufficiently large $D$,
the evolution of $\p2$ is dominated by the diffusive term $2 Dt$ in the early stages.
This allowed us to deduce that the Zeno effect is suppressed on a timescale
\beq
\tau = \frac { m \hbar } { D \eps} = \frac { \hbar^2 } { D a L }.
\eeq
This may be interpreted as the decoherence time on the lengthscale
\beq
(aL)^{\half} = \left( \frac { \hbar \eps } { m } \right)^{\half},
\eeq
which is the effective spatial grid size associated with the projections a time $\eps$
apart.

We showed, mainly from numerical analysis, that most initial states go into a stationary
regime and we deduced the characteristic scales of that regime. We argued that there
would be no Zeno effect in that regime as long
as the parameters of the model are such that the final energy time satisfies
$t_E^f \ll \eps$.

Finally, in Section X, we illustrated the Zeno effect and its suppression by decoherence using
diagrams showing the probability flux field in space and time. These gave a very
clear visual impression as to how the Zeno effect and its suppression work. (They
are equivalent to Bohm trajectories when there is no environment present).
They show very clearly both the Zeno and anti-Zeno effects, with the Zeno effect
portrayed as bunching up of the trajectories towards the centre of the region, to
a degree sufficient to exceed the anti-Zeno effect at the boundary produced by the
large momentum spreading there. The environment produces a smoothing of the trajectories
wiping out both the Zeno and anti-Zeno effects and restoring the expected classical
escape.

In all these calculations, we assumed an initial state with $ \langle p \rangle =
0$. This is appropriate since the case of an initially moving packet is very closely
related to the situation of reflection off a single step potential which was analyzed
in detail in our earlier, closely related paper, Ref.\cite{BeHa}. Furthermore, as
noted in the Introduction, the large fluctuations problem encountered in that earlier
work are not problematic here, since we are working with a finite region $[-L/2,L/2]$.
This means that even if very large fluctuations are required to kill the quantum
effects, these fluctuations enhance the escape from the region, thereby contributing
to the restoration of classical behaviour.

In conclusion, we have established the conditions and timescale under which the Zeno
effect relating to escape from a spatial region is suppressed thereby restoring a sensible classical
limit. Related issues will be analyzed further in future publications.

\section{Acknowledgments}

We are grateful to James Yearsley for useful conversation. This work was supported
by EPSRC grant No. EP/J008060/1.

\bibliography{apssamp}

\end{document}